\documentclass[authoryear,review,3p]{elsarticle}

\journal{arXiv}

\usepackage{lineno}
\usepackage[utf8]{inputenc}
\usepackage{csquotes}
\usepackage{amsmath}
\usepackage{amssymb}
\usepackage{color}
\usepackage{booktabs,multirow}
\usepackage{rotating}
\usepackage{array}
\usepackage{rotating}
\usepackage{longtable}
\usepackage{mathabx}
\usepackage{paralist}
\usepackage[inline]{enumitem}
\usepackage{bibentry}
\usepackage[dvipsnames]{xcolor}

\DeclareMathOperator*{\argmax}{arg\,max}
\DeclareMathOperator*{\argmin}{arg\,min}

\begin{document}

\begin{frontmatter}

\title{Micro-level dynamics in hidden action situations with limited information\tnoteref{mytitlenote}}
\tnotetext[mytitlenote]{This work is supported by the funds of the \"Osterreichische Nationalbank [Austrian Central Bank Anniversary Fund, Grant No. 17930].}

\author[klagenfurt]{Stephan Leitner\corref{mycorrespondingauthor}}
\cortext[mycorrespondingauthor]{Corresponding author.}
\ead{stephan.leitner@aau.at}
\author[klagenfurt]{Friederike Wall}
\ead{friederike.wall@aau.at}
\address[klagenfurt]{University of Klagenfurt, Universit\"atsstra{\ss}e 65-67, 9020 Klagenfurt, Austria}

\begin{abstract}
The hidden-action model provides an optimal sharing rule for situations in which a principal assigns a task to an agent who makes an effort to carry out the task assigned to him. However, the principal can only observe the task outcome but not the agent's actual action\textcolor{black}{, which is why the sharing rule can only be based on the outcome}. The hidden-action model builds on somewhat idealized assumptions about the principal's and the agent's capabilities related to information access. We propose an agent-based model that relaxes some of these assumptions. Our analysis lays particular focus on the micro-level dynamics triggered by limited \textcolor{black}{access to} information. For the principal's sphere, we identify the so-called \textit{Sisyphus effect} that explains why the sharing rule \textcolor{black}{that provides the agent with incentives to take optimal action is difficult to achieve} if the information is limited, and we identify factors that moderate this effect. In addition, we analyze the behavioral dynamics in the agent's sphere. We show that the agent might make even more of an effort than optimal under unlimited \textcolor{black}{access to} information, which we refer to as \textit{excess effort}. Interestingly, the principal can control the probability of making an excess effort via the incentive mechanism. However, how much excess effort the agent finally makes is out of the principal's direct control. 
\end{abstract}

\end{frontmatter}

\modulolinenumbers[5]

\section{Introduction}

Researchers in management and economics have already recognized that the assumptions included in some of the models employed in these fields are somewhat idealized and often do not reflect the characteristics of real-world decision-makers \citep{Kohn2004,Axtell2007}. While, of course, these models are technically correct and valid, due to the assumptions they build on, they may lack the power to explain empirical phenomena \citep{Franco2020,Leitner2014,Wall2020,Roberts2012,Lambright2009}. \cite{Axtell2007} refers to the core of the idealized assumptions as the \textit{neoclassical sweet spot}, which includes rationality, agent homogeneity, equilibrium, and non-interactiveness. Building on this sweet spot usually increases the models' internal validity, but this often comes at the cost of external validity, which, in consequence, sometimes results in a focus on problems of little substantive interest for corporate practice \citep{Shapiro2005,Cohen2006}. Also, in the context of mechanism design, it has been recognized that assumptions around the neoclassical sweet spot might be overly restrictive since decision-makers might make errors due to bounded rationality. For the \enquote*{mechanism designer} such errors are unpredictable and might unfold adverse effects if not adequately considered \citep{Halpern2008,Wright2012,Liu2016}. Recently, calls for bridging disciplines and including findings from related fields -- such as cognitive psychology -- into economic models have kept emerging due to the identified problems \citep{Royston2013,Wall2020,Hamalainen2013}. For example, previous research already started to account for somewhat \enquote*{incompetent} agents in the context of principal-agent models and suggests compensating for it, for example, by additional guidance \citep{Hendry2002}. However, the focus is mainly put on limitations in the the agent's sphere, while for the principal, the assumptions in the neoclassical sweet spot are often perpetuated. 

We recognize the raised concerns and take up on this previous line of research by proposing an agent-based model of the hidden-action problem in which the principal\textcolor{black}{'s} and the agent\textcolor{black}{'s information} are simultaneously limited. \textcolor{black}{In particular, we adopt the idea of \textit{limited} information introduced in \cite{Frieden2010} and \cite{Hawkins2010}: Suppose the information \textit{intrinsic to} a particular system is \(J\). Systems could be entire organizations or the environment in which an organization resides, and information \(J\) represents the most complete and perfectly knowledgeable information concerning this particular system. We denote the information \textit{about} the same system by \(I\), whereby \(I\) is, e.g., based on observations and learning. Any gathering of information is represented by the information flow process \(J \rightarrow I\). Consequently, the distance \(I - J\) indicates how easily one is informed about the system or how well one can access information \(J\).}
\textcolor{black}{The proposed agent-based model is constructed following the approach of agentization \citep{Guerrero2011,Leitner2014}.}%

Holmstr\"om's \textcolor{black}{hidden-action} model captures a principal who assigns a task to an agent. The agent acts on behalf of the principal by making an effort to carry out the assigned task. The principal’s role is to provide capital and incentives. Information asymmetries further characterize the relationship between the principal and the agent. The principal can only observe the task outcome but not the actual state of the environment and the action taken by the agent -- it is \textit{hidden} --, which is why the principal can only employ a performance-based compensation mechanism. All other pieces of information are assumed to be observable by the principal and the agent. \textcolor{black}{The model provides an optimal incentive mechanism, i.e., a rule to share the task outcome between the principal and the agent (that includes optimal risk-sharing) \citep{Holmstrom1979,Lambert2001,Eisenhardt1989}.}
\textcolor{black}{Among the assumptions about information that are specifically addressed in the agent-based model is the one that the principal and the agent} are perfectly informed about the distribution of the variable representing the environment and the one that they have full access to all feasible ways of carrying out the task at hand.\footnote{Further assumptions include, for example, that the principal is fully aware of the agent's characteristics in terms of the utility function, productivity, and reservation utility. The reader is referred to \cite{Lambert2001} and \cite{Eisenhardt1989} for reviews on the principal-agent literature.} 

The structure of Holmstr\"om's hidden-action model can capture a multiplicity of real-world delegation relationships, for example, between employer and employee, buyer and supplier, or homeowner and contractor, and, therefore, it appears evident that it has implications for how incentive mechanisms are designed in practice \citep{Caillaud2000,Leitner2020,Bhattacharya2019}. However, empirical evidence supports the conjecture that the assumptions included in the hidden-action model are not necessarily in line with the characteristics of real-world decision-makers \citep{Eisenhardt1989,Hendry2002}. This makes it particularly dangerous to employ the outcomes of the hidden-action model in real-world settings without having substantive knowledge of the dynamics that might be triggered by violating the assumptions in the neoclassical sweet spot. This gap is where we place our research. We aim at (better) understanding the consequences of limited \textcolor{black}{access to} information about the environment and the action space in hidden-action setups and want to answer the following questions: How does
\begin{enumerate*}[label=\textit{(\roman*)}]
\item limited \textcolor{black}{access to} information for the principal and the agent affect individual behavior and micro-level dynamics in the context of hidden-action problems, and 
\item what are the macro-level patterns that emerge from these micro-level dynamics?
\end{enumerate*}

The remainder of this paper is structured as follows: Section \ref{sec:2} \textcolor{black}{introduces Holmstr\"om's hidden-action model and discusses related research.} 
In Sec. \ref{sec:abm}, we transfer \textcolor{black}{the hidden-action problem} into an agent-based model following the method of agentization. The results and a discussion are included in Sec. \ref{sec:results}, where we present the Sisyphus effect and excess effort to describe the dynamics in the principal's and the agent's sphere, respectively, and discuss the 
results. Finally, Sec. \ref{sec:conclusion} concludes the paper. 

\section{Holmstr\"om's hidden-action model and related research}
\label{sec:2}

\subsection{\textcolor{black}{Holmstr\"om's hidden-action model}}
\label{sec:holmstrom}

\begin{figure}
\center
  \includegraphics[width=0.7\linewidth]{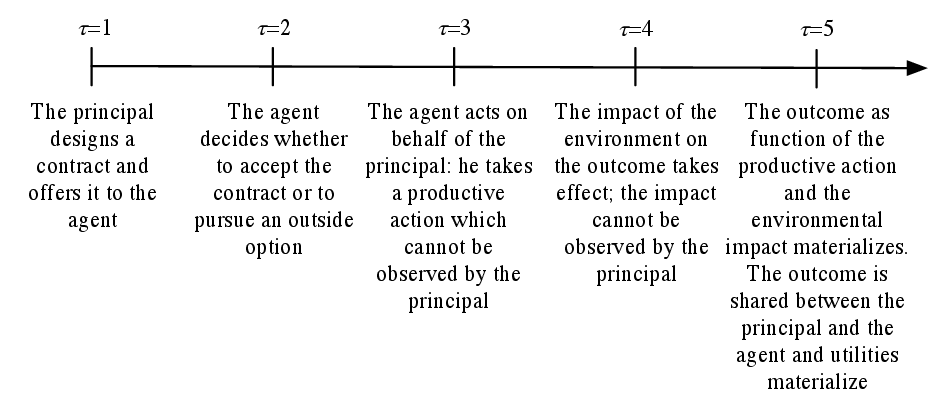}
  \caption{Sequence of events in Holmstr\"om's hidden-action model}
  \label{fig:standard-hidden-action}
\end{figure}

\paragraph{\textcolor{black}{The hidden-action model}}
One of the first versions of the hidden-action model was introduced in \cite{Holmstrom1979}. This model captures the relationship between a risk-neutral principal  who assigns a task to a risk-averse agent.\footnote{Further model variants are, for example, introduced in \cite{Harris1979} and \cite{Spence1971}. Extensive reviews of the principal-agent literature are provided in \cite{Lambert2001} and \cite{Eisenhardt1989}.} The principal's role is to provide capital and construct incentives, while the agent's role is to act on behalf of the principal. 
The sequence of events within the hidden-action model is illustrated in Fig. \ref{fig:standard-hidden-action}: 
The principal offers the agent a contract in \(\tau=1\). This contract includes a task to be carried out and a rule used to share the outcome associated with the task. If the agent accepts the contract in \(\tau=2\), he takes productive action \(a\in\mathbf{A}\subseteq\mathbb{R}^{\textcolor{black}{+}}\) to carry out the task in \(\tau=3\). The action \(a\) cannot be observed (at reasonable costs) by the principal; it is \textit{hidden}. Together with a random exogenous variable \(\theta\), the taken action \(a\) leads to an outcome according to the production function \(x=x(a,\theta)\). The principal cannot observe \(\theta\). The agent, however, has information about \(\theta\) after it has taken effect in \(\tau=4\).

The hidden-action model proposes a rule \(s(\cdot)\) to share the outcome \(x\) between the principal and the agent. This rule assures that the agent accepts the contract in \(\tau=2\) (participation constraint) and that he has incentives to take the productive action that is value-maximizing from the principal's point of view (incentive compatibility constraint). 

The agent's share of \(x\) is denoted by \(s(x)\). The principal is risk-neutral and experiences utility from \(x-s(x)\). She wants to maximize her utility 
\begin{equation}
U_P(x-s(x)) = x - s(x)~.
\end{equation}
The risk-averse agent experiences utility \(V(s(x))\) from  his share of \(x\) and disutility \(G(a)\) from taking action \(a\).\footnote{Where $V^\prime > 0$ and $x_a \geq 0$. Subscript $a$ denotes the partial derivative with respect to a.} He wants to maximize his utility function
\begin{equation}
U_A(s(x),a) = V(s(x)) - G(a)~. 
\end{equation}
In consequence, the principal faces the following optimization problem: 
\begin{subequations}
\begin{align}
\max_{s(\cdot),a} 	\quad 	& 	{E}\left(U_P\left(x-s\left(x\right)\right)\right) \label{eq1:maximization}\\
\textrm{s.t.} 	\quad 	&	{E}\left(U_A\left(s\left(x\right),a\right)\right)\geq\underline{U} \label{eq1:PC}\\
 					 &	a \in \argmax_{a^\prime \in \mathbf{A}}  {E} \left(U_A\left(s\left(x\right),a^\prime\right)\right)~. \label{eq1:ICC}
\end{align}
\end{subequations}
\(\underline{U}\) stands for the agent's utility gained from the outside option. Equations \ref{eq1:PC} and \ref{eq1:ICC} represent the participation and the incentive compatibility constraint, respectively. The operator \enquote*{arg max} maximizes the set of arguments that maximize the objective function that follows in the equation. \textcolor{black}{A solution to the program that is formalized in Eqs. \ref{eq1:maximization} to \ref{eq1:ICC} is included in \ref{app:a}.} 

\paragraph{\textcolor{black}{Information access in Holmstr\"om's hidden-action model}}
A vital feature of Holmstr\"om's hidden-action model lies in the principal's and the agent's respective access to information: They share the same information about \textit{(i)} the distribution of $\theta$ and \textit{(ii)} the action space $\mathbf{A}$. In addition, both parties are fully aware of \textit{(iii)} the production function and can observe \textit{(iv)} the outcome \textit{after} it has been realized in \(\tau=5\). The principal is informed about \textit{(v)} the agent's characteristics (e.g., productivity, utility function, and utility the agent can experience from an outside option). The hidden-action model introduced in \cite{Holmstrom1979} assumes that both the principal and the agent are endowed with the required means and capabilities to access information \textit{(i)-(v)}.

The principal, however, does \textit{not} have access to \textit{(vi)} the action taken by the agent in \(\tau=3\) and \textit{(vii)} the actual state of the environment, which takes effect in \(\tau=4\).\footnote{The principal, at least, cannot observe the action taken by the agent at a reasonable cost.} Information \textit{(vi)} and \textit{(vii)} are \textit{private} for the agent, and the principal cannot infer the agent's personal information from the observable outcome measure. Therefore, the sharing rule proposed by this framework cannot be based on the action taken by the agent but is based on the outcome alone. The assumptions about information access included in the hidden-action model are summarized in Tab. \ref{tab:info-access}.

\subsection{\textcolor{black}{Related research on assumptions in Holmstr\"om's hidden-action model}}

\textcolor{black}{The related literature on relaxing assumptions in the principal-agent framework can mainly be divided into a positive and a normative stream \citep{Shapiro2005,Eisenhardt1989}. The positive principal-agent literature is primarily empirical, whereas the normative stream usually follows a mathematical (and non-empirical) approach. Most of the related work on relaxing assumptions related to the hidden-action model is part of the positive stream because relaxing assumptions usually goes hand in hand with a decrease in the mathematical tractability of models. In this vein, \cite{latacz1997} and \cite{Holmes2010,holmes2017} argue that it is difficult for theory to provide closed-form solutions to real-world problems because modelling a real-word setting requires relaxed and simplified assumptions, as opposed to existing mathematical models. In the context of relational contracts, \cite{Gil2016} argue that, amongst others, the assumption of symmetric information must be relaxed in the existing literature to create more empirically sound models. \cite{englmaier2016} add that this might lead to interesting dynamics such as non-stationary optimal contracts and an analysis of how such agreements evolve. This context is where we place our research: Agent-based modelling allows us to develop models that feature relaxed assumptions. These models are then \enquote*{solved} by numerical simulation, and the emerging behavioral patterns, as well as the corresponding driving forces, are analyzed.}

\textcolor{black}{
In a related stream of research, \cite{Keser2007} are concerned with theories of the agent's behavior in hidden-action setups. They carry out laboratory experiments and analyze how well their fair-offer theory \citep{keser2000} explains the agent's behavior compared to the standard principal-agent theory.\footnote{the fair-offer theory suggests that the principals provides the agent with full insurance against losses, and additionaly the agent receives at most $50\%$ of the net surplus \cite{keser2000}.} They find that the fair-offer theory explains the agent's behavior accurately. Still, the predictive power decreases with decreasing net surplus. \cite{hoppe2018} are concerned with the contractibility of the outcome. They perform laboratory experiments and test which contracts are negotiated if the outcome is either contractible or not. They find that in most cases, incentive-compatible contracts are negotiated in the case of a contractible outcome. However, in cases with a non-contractible outcome, a relatively low effort is chosen in most cases. \cite{bisin2004} are concerned with the assumption of exclusive contracts, i.e., one party in an agreement can restrict the other party from engaging from contractual relationships with other agents. From an information perspective, this assumption requires that the principal can ideally monitor the agent's contracts with other institutions. \cite{bisin2004} follow a formal approach and characterize equilibria in such situations; in particular, they show that agents enter multiple contractual relationships and intermediaries make profits in equilibrium. \cite{iossa2015} are concerned with general information structures in principal-agent models. They argue that a more complete way of modelling principal-agent relationships requires making information structures endogenous. They follow a formal approach and provide a model that features agents who invest in private information gathering and show that -- depending on the agent's endogenous decision about investing in information -- situations with hidden-action and hidden-information emerge with positive probability. }

\textcolor{black}{
Recently, \cite{Reinwald2020,reinwald2021,reinwald2022} have analyzed situations in which both the principal and the agent have incomplete information about the environment, and are characterized by limited and asymmetric information processing capabilities. They find that the agent's utility is over-dependent on the principal's choices, so even if agents invest in information gathering, their utility barely increases.
In a recent paper, \cite{Leitner2020} are also concerned with the information assumptions in hidden-action setups. They analyze the emerging organizational performance in the case of principal and agent employing different information systems. While \cite{Reinwald2020,reinwald2021,reinwald2022} exclusively focus on limitations in information about the environmental variable, \cite{Leitner2020} also consider limitations in information about the action space. However, their analysis almost exclusively focuses on the organizational level. We move forward with this line of research and provide an in-depth analysis of the behavioral micro-level dynamics in hidden-action situations that are induced by limited information access. 
}

\section{An agent-based model of hidden-action situations with limited information}
\label{sec:abm}

The following problem is transferred from the hidden-action model introduced in Sec. \ref{sec:holmstrom} to an agent-based model (ABM): The \textcolor{black}{risk-neutral} principal offers the \textcolor{black}{risk-averse} agent a contract that, amongst others, includes a specific task to be carried out and a \textcolor{black}{linear} rule to share the task outcome. The agent accepts the contract and takes a productive action on behalf of the principal, which, together with an exogenous factor, produces the (monetary) outcome. The principal experiences utility gained from her share of the outcome, while the agent's utility results from his share of the outcome minus the disutility gained from taking action. To assure that the agent takes the action that maximizes the principal's utility, the principal aims at finding the optimal \textcolor{black}{parameters for the} incentive scheme. 

Compared to the hidden-action model introduced in Sec. \ref{sec:holmstrom}, the ABM includes \textit{less} restrictive assumptions about the principal's and the agent's respective access to information. We put particular emphasis on the accessibility of information about the environment and on access to information about the action space.\footnote{In line with the hidden-action model introduced in Sec. \ref{sec:holmstrom}, the principal and the agent share information about the outcome and the agent's characteristics in the agent-based model.} Recall that the accessibility of information assumed in the model introduced in Sec. \ref{sec:holmstrom} allows the principal to find the optimal \textcolor{black}{sharing rule} within only one sequence of events summarized in Fig. \ref{fig:standard-hidden-action}. The ABM, in contrast, captures situations in which the access to relevant information is limited, which is why the principal can no longer find the optimal solution immediately. Consequently, one distinctive feature of the model introduced here is that the principal is required to \textit{search} for the optimal \textcolor{black}{parameters for the linear sharing rule} over time by employing a hill-climbing-based search algorithm.\footnote{The model is implemented using MathWorks\textsuperscript{\textregistered} Matlab (R2021B).} \textcolor{black}{Holmstr\"om's hidden action model is rather generic in its formulation. Translating this model into an agent-based model and \enquote*{solving} it via numerical simulations requires specifying some model elements, such as the utility and production functions, in more detail. In particular, the agent-based model features a linear sharing rule, and the principal can adapt the premium parameter of this rule. In contrast to our model, the form of this rule is not known beforehand in Holmstr\"om's model. We introduce the translated model in the following subsection, and Tab. \ref{tab:comparison} compares the main model elements as specified in Holmstr\"om's hidden-action model and the agent-based version. }

\begin{table}
\textcolor{black}{
\begin{tabular}{p{0.3\textwidth} p{0.3\textwidth}  p{0.3\textwidth}  }
\noalign{\smallskip}\hline \noalign{\smallskip} 
Function & Holmstr\"om's model & Agent-based model 	\\
\hline\noalign{\smallskip} 
Principal's utility     & $U_P (x - s(x))$          & $U_P = x_t - s(x_t))$         \\
Agent's utility         & $U_A = V(s(x)) - G(a)$    & $U_A = \frac{1-e^{-\eta\cdot s(x_t)}}{\eta} - 0.1a_{{t}}^2$  \\
Production function     & $x=x(a, \theta)$          & $x_t = a_t \cdot \rho + \theta_t$   \\
Sharing rule   & $s(x) $                   & $s(x_t) = x_t \cdot \phi_t$   \\
\hline\noalign{\smallskip} 
\end{tabular}
\caption{Comparison of the main model elements in Holmstr\"om's model and the agent-based version thereof}
\label{tab:comparison}}
\end{table}

\subsection{Process overview and scheduling}
\label{sec:model-overview}

\begin{figure}
\center
  \includegraphics[width=1\linewidth]{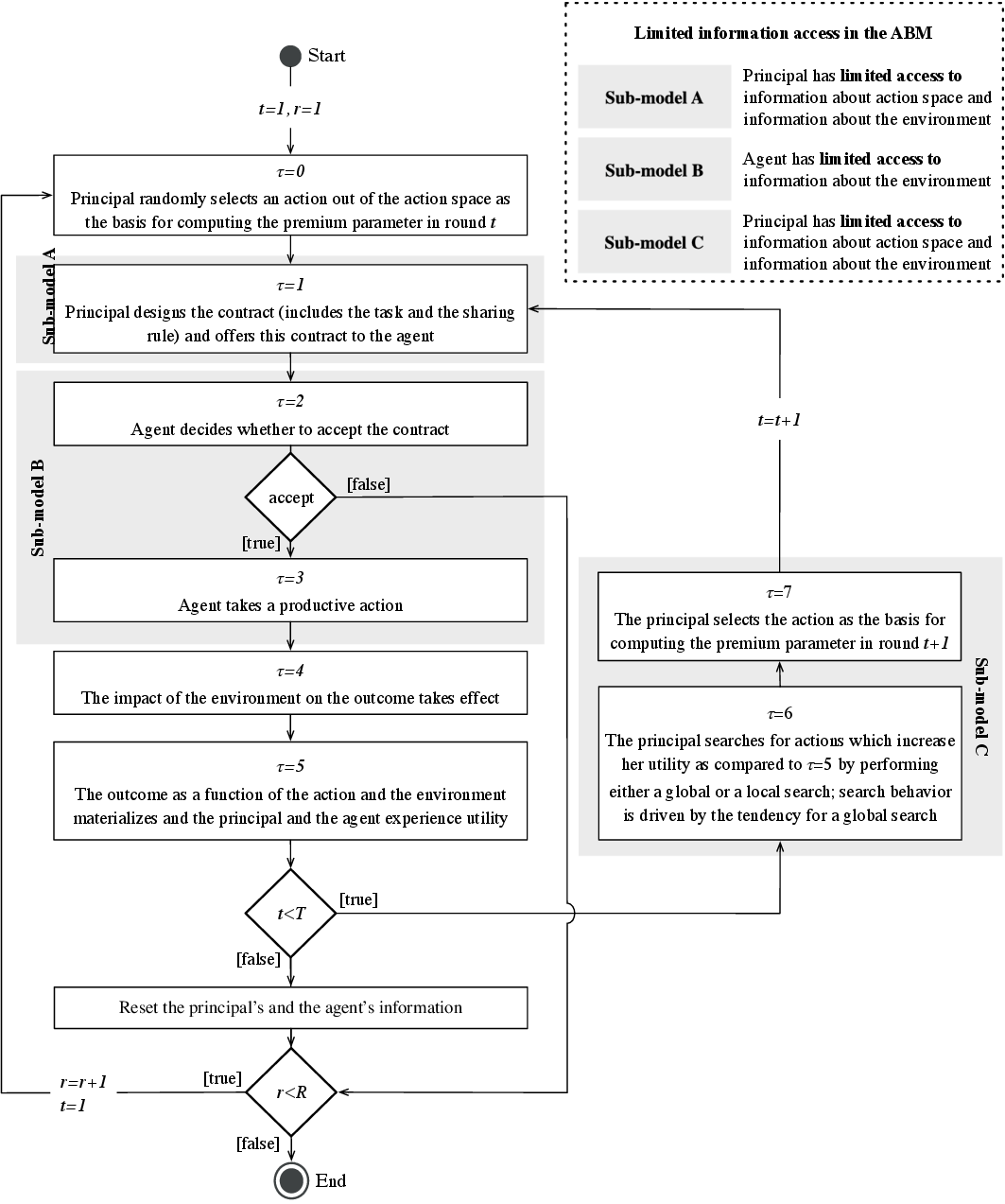}
  \caption{Process overview}
  \label{fig:abm}
\end{figure}

The flowchart in Fig. \ref{fig:abm} gives an overview of the process implemented in the ABM. We distinguish between three indices: First, \(r\) \textcolor{black}{$\in$} \(\{1,...,R\} \subset \mathbb{N}\) stands for the number of simulation runs. Second, \(t\) \textcolor{black}{$\in$} \(\{1,...,T\} \subset \mathbb{N}\) stands for rounds (i.e., periods) within one simulation run and, third, \(\tau\) \textcolor{black}{$\in$} \(\{1,...,7\} \subset \mathbb{N}\) indicates the sequence of events within one period. In the initial period \(t=1\), the principal starts \textcolor{black}{with a random parameter for the compensation function}. She designs the contract \textcolor{black}{and offers it to the agent in \(\tau=1\)}. The agent decides whether to accept the contract or not and, if he accepts, takes a productive action in \(\tau=2\) and \(\tau=3\), respectively. The environment takes effect in \(\tau=4\), while in \(\tau=5\) the outcome materializes, and the principal and the agent experience utility. There is a chance that the initial \textcolor{black}{parameter of the sharing rule} is not the optimal one, \textcolor{black}{i.e., it might not provide the agent with incentives to take the optimal action}. The principal is aware of the existence of further \textcolor{black}{ways to carry out the task} in the action space that might result in a higher utility for her, and, in periods \(t=2,\dots,T\) she searches for \textcolor{black}{these actions}. In particular, she does so in \(\tau=6\), and in \(\tau=7\), and she selects \textcolor{black}{an action to construct the incentives} in the next period \(t+1\). \textcolor{black}{Please note that in Holmstr\"om's hidden action model, the principal comes up with a sharing rule that provides the agent with incentives to make the optimal effort, i.e., to take the utility-maximizing action. However, in the proposed agent-based model, the principal is not necessarily aware of the optimal action. Rather, she selects the action that appears to be utility-maximizing, given the information available to her at a particular point in time, and then comes up with a parameter for the linear sharing rule that provides the agent with incentives to take that particular action.} The model moves to period $t+1$ after this decision, and the principal
designs a contract again in \(\tau=1\).\footnote{Please note that the principal searches for action that promise to increase her utility, and then she designs the sharing rule so that the agent has incentives to take these actions. Throughout the paper, we sometimes take a shortcut and argue that the principal searches for the optimal incentive scheme (which includes both searching for the best possible action \textcolor{black}{given the information available to her at a particular point in time} and designing \textcolor{black}{the corresponding} sharing rule).  } In the course of the contractual relation, the principal and the agent collect and learn specific pieces of information. The entire sequence is repeated \(R\) times. 
The contract design is captured in \textit{sub-model A} \textcolor{black}{(Sec. \ref{sec:submodel-a})}, details about the agent's decision related to his productive action are provided in \textit{sub-model B} \textcolor{black}{(Sec. \ref{sec:submodel-b})}, and the principal's hill-climbing-based search is put in concrete terms in \textit{sub-model C} \textcolor{black}{(Sec. \ref{sec:submodel-c})}. 
\subsection{\textcolor{black}{The principal's and the agent's respective characteristics}} 

\subsubsection{The principal's \textcolor{black}{characteristics}}
The principal's main roles are to provide capital and incentives. She is risk-neutral and aims at maximizing the utility gained from her share of the outcome. 
We formalize her utility function by 
\begin{equation}
U_P(x_t - s(x_t)) = x_t - s(x_t)~. 
\label{eq:principal}
\end{equation}
The outcome in \(t\) is denoted by 

\begin{equation}
x_t = a_t \cdot \rho + \theta_t~,
\label{eq:prod-function}
\end{equation}
\noindent where \(a_t\) \(\textcolor{black}{\in \mathbb{R}^{+}}\) stands for the action taken by the agent in \(t\)\textcolor{black}{,} \(\rho \in [0,1]\) indicates the agent's productivity, \textcolor{black}{and} \(s(\cdot)\) represents the sharing rule. \textcolor{black}{The {environment} is modeled as a single random variable $\theta$ that follows a Normal Distribution, $\theta_t \sim N \left(\mu, \sigma \right)$. It captures, for example, actions of competitors and the government that affect the task outcome.} 
The agent's share of the outcome follows the linear function 
\begin{equation}
\label{eq:sharing-rule}
 s(x_t)=x_t \cdot \phi_t   ~,
\end{equation}
\noindent where \(\phi_t \in [0,1]\) stands for the premium parameter effective in period $t$. Note that the premium parameter has to be specified by the principal in advance in \(\tau=1\) and cannot be changed \textit{within} one period. However, \textit{between} periods (from $t$ to $t+1$), the principal can adapt it (see \textit{sub-models A and C}).\footnote{Please note that the sharing rule is the principal's only means to control the agent's behavior: The principal \textcolor{black}{searches for} an action in \textit{sub-model C} and \textcolor{black}{constructs the incentives} that provide the agent with incentives to take \textcolor{black}{this particular} action in \textit{sub-model A}.} 

\subsubsection{The agent's \textcolor{black}{characteristics}}
The agent's role is to take productive action to perform the task assigned to him. He is risk-averse and aims at maximizing the utility gained from his share of the outcome minus the disutility gained from taking action. We formalize the agent's CARA utility function by 
\begin{equation}
U_A(s(x_t),a_t) = \frac{1-e^{-\eta\cdot s(x_t)}}{\eta} - 0.1a_{\textcolor{black}{t}}^2~,
\label{eq:agent}
\end{equation}
\noindent where \(\eta\in\mathbb{R}\) stands for the agent's Arrow-Pratt measure of risk-aversion. The agent can influence his disutility via the choice of the productive action taken in \(\tau=3\), which affects the produced outcome. The agent's decision as to which action to take is included in \textit{sub-model B}. In addition, the agent is characterized by a reservation utility $\underline{U}$ and his productivity $\rho$. The former is the utility the agent could experience from an outside option, while the latter defines the relative contribution of a productive action to the outcome.

\begin{table}
\centering
\begin{tabular}{p{0.05\textwidth} >{\raggedright\arraybackslash}p{0.25\textwidth} >{\centering\arraybackslash}p{0.12\textwidth} >{\centering\arraybackslash}p{0.12\textwidth} >{\centering\arraybackslash}p{0.12\textwidth} >{\centering\arraybackslash}p{0.12\textwidth}}
\multicolumn{2}{l}{ \multirow{2}{*}{\textbf{Types of information}} }  & 	\multicolumn{2}{c}{\textbf{Hidden-action model\(^{a}\)} }  	&  \multicolumn{2}{c}{\textbf{ABM\(^{b}\)} }  \\
\noalign{\smallskip} \cline{3-6}\noalign{\smallskip} 
&										& 	Principal	&	Agent	& 	Principal	&	Agent	\\
\hline\noalign{\smallskip} 
\textit{(i)}	&	Distribution of the exog. factor					&	full  		&	full 		&	no  			&	no 		\\ \noalign{\smallskip} 
\textit{(ii)}	&	Action space												&	full  		&	full 		&	lim	  		&	full 		\\ \noalign{\smallskip} 
\textit{(iii)}	&	Production function											&	full  		&	full 		&	full  			&	full 		\\ \noalign{\smallskip} 
\textit{(iv)}	&	Outcome (after it has materialized)						&	full  		&	full 		&	full  			&	full 		\\ \noalign{\smallskip} 
\textit{(v)}	&	Agent's characteristics 										&	full  		&	full 		&	full  			&	full 		\\ \noalign{\smallskip} 
\textit{(vi)}	&	Agent's action										&	no  		&	full 		&	no  			&	full 		\\ \noalign{\smallskip} 
\textit{(vii)}	&	Exogenous variable\(^{c}\) 								&	no  		&	full 		&	no	  		&	full 		\\ \noalign{\smallskip} 
\textit{(viii)}  &	 Estimations and observations of the exogenous variable						& 	n.a.		&	n.a.		&	lim/full		&	lim/full		\\ \noalign{\smallskip} 
\hline\noalign{\smallskip} 
\end{tabular}
\begin{flushleft}\footnotesize{
Access to information: full=full access, lim=limited access, lim/full=limited or full access (depending on parameterization), no=no access, n.a.=not available in the model. \\
\(^{a}\)Hidden-action model, see Sec. \ref{sec:holmstrom}. \\ 
\(^{b}\)Agent-based model, see Sec. \ref{sec:abm}.\\
\(^{c}\)Only the agent has this information. In the ABM, the principal \textit{estimates} the exogenous variable.
}\end{flushleft}
\caption{Information in the hidden-action model and the agent-based model }
\label{tab:info-access}
\end{table}

\subsection{The principal's and the agent's respective limited information}
\label{sec:p-and-a-info-access}

\subsubsection{The principal's limitations in information}
The principal has limited access to
\begin{enumerate*}[label=\textit{(\roman*)}]
\item information about the exogenous variable $\theta_t$, and 
\item information about the action space the agent can choose from when carrying out the task. 
\end{enumerate*}
The assumptions related to the principal's access to information are summarized in Tab. \ref{tab:info-access}.

The principal cannot observe \textit{(i) the exogenous variable} $\theta_t$ after it has taken effect in \(\tau=4\) and has no information about its distribution. However, she is endowed with the capabilities to \textit{estimate} the exogenous variable $\theta_t$ and store the estimation in her memory.
Recall, in every period $t$ the agent performs the action \(a_t\) in \(\tau=3\), the environment $\theta_t$ takes effect in \(\tau=4\), and the outcome \(x_t=a_t \cdot \rho + \theta_t\) materializes in \(\tau=5\) (see Fig. \ref{fig:abm}). %
The principal can observe the outcome \(x_t\) after it has materialized in \(\tau=5\) and knows the action \(\tilde{a}_t\) that \textcolor{black}{was the basis for fixing the incentive parameter in} \(\tau=3\) (for details about \(\tilde{a}_t\) see \textit{sub-model A}). The principal estimates the exogenous variable in \(t\) according to 
\begin{equation}
{\vartheta}_t = x_t - \tilde{a}_t \cdot \rho~,
\end{equation}
\noindent and stores the estimation in her memory.
We denote the previous estimations accessible to the principal in period $t$ by
\begin{equation}
    {\mathbf{M}}_{Pt}= [{\vartheta}_{t-1}, \dots, {\vartheta}_{t-m}]~,
\end{equation}
where the parameter \(m\) defines the principal's memory in terms of the number of previously estimated exogenous variables accessible to her so that the length of \(\mathbf{M}_{Pt}\) is equal to $m$.\footnote{Please note that if the principal made less than $m$ estimations (in periods $t<m$), she only has those $t$ number of estimations available.} The principal's prediction of the environment in \(t\) is computed as the mean $\mu(\cdot)$ of the accessible estimations so that 
\begin{equation}
E_P(\theta_t) = \mu (\mathbf{M}_{Pt} )~.
\label{eq:theta}
\end{equation} 
The principal is also limited in her access to \textit{(ii) information about the action space}. We denote the feasible actions \(t\) from the principal's  point of view by \(\mathbf{A}_{Pt}\), and define the lower and the upper boundaries by the participation and the incentive compatibility constraint, respectively \citep{Holmstrom1979,Leitner2020}.\footnote{Actions within the action space are ordered from smallest to largest.} Recall that computing the two constraints includes the outcome. Since  the environment contributes to the outcome, we use the principal's prediction of the environment and reformulate the production function as  
\begin{equation}
\tilde{x}_{Pt} = a_t \cdot \rho + E_P(\theta_t)~.
\label{eq:exp-outcome-principal}
\end{equation} 
\noindent Then, we can formalize the lower boundary \(\underline{a}_{Pt}\) as the smallest feasible element of \(\mathbf{A}_{Pt}\) that fulfills the participation constraint, \textcolor{black}{$\underline{a}_{Pt} = \argmin_{a{'}\in \mathbf{A}_{Pt}} \{a{'}: U_A(s(\tilde{x}_{Pt}, a{'})) \geq \underline{U}\} $}.
%
%
Corresponding to the incentive compatibility constraint, we define the upper boundary as the largest feasible action and formalize it by 
\(\overline{a}_{Pt} = \argmax_{a^{\prime} \in \mathbf{A}_{Pt}} U_A (s(\tilde{x}_{Pt}),a^{\prime}  ))\). We refer to $\mathbf{A}_{Pt}$ as the principal's \textit{predicted action space} in period $t$; \textcolor{black}{and it is the basis for the formulation of the principal's search spaces.}

The principal accesses $\mathbf{A}_{Pt}$ when she searches for alternative actions that promise a higher utility than the action recently taken by the agent. To do so, the principal can either perform a local or a global search. 
The former results in \textcolor{black}{a search space that includes actions that are} \textit{similar} to the \textcolor{black}{recent} action $\textcolor{black}{\tilde{a}_t}$, while the latter means that the principal has access to \textcolor{black}{a search space that includes} actions that are more \textit{different} from the recent action. \textcolor{black}{The boundaries of the search spaces are} controlled by parameter \(\lambda\), whereby \(1/\lambda\) and \(1-1/\lambda\) determine the fractions of the action space accessible in case of a local and global search, respectively. Details on the search procedure are provided in \textit{sub-model C}. 

\subsubsection{The agent's limitations in information} 
While the assumption of the agent's full accessibility of information related to the action space is transferred from the hidden-action model introduced in Sec. \ref{sec:holmstrom}, the ABM assumes that the agent has limited access to information about the environment. However, unlike the principal, the agent can \textit{observe} the environment's state after it has taken effect. We denote the observations of the exogenous variable available to the agent in period $t$ by\footnote{If the agent makes less than $m$ observations (in periods $t<m$), he only has those estimations available.}
\begin{equation}
    \mathbf{M}_{At}= [\theta_{t-1}, \dots, \theta_{t-m}]~.
\end{equation}
Parameter $m$ controls the agent's memory. The agent's prediction of the environment in \(t\) is also computed as the mean $\mu(\cdot)$ of the accessible estimations so that 
\begin{equation}
E_A(\theta_t) = \mu (\mathbf{M}_{At} )~.
\label{eq:theta_abm}
\end{equation} 

\noindent \textcolor{black}{
As is the case for the principal, the agent's prediction of the environment in period $t$ affects the agent's expected outcome in $t$: 
\begin{equation}
\label{eq:expected-outcome-agent}
\tilde{x}_{At} = a_t \cdot \rho + E_A(\theta_t)~.
\end{equation} }

\subsection{Sub-models}
\label{sec:sub-models}

\subsubsection{Sub-model A}
\label{sec:submodel-a}
The first sub-model covers the principal's decision about the premium parameter $\phi_t$ in period $t$. The input to this sub-model is the action \textcolor{black}{that is utility maximizing from the principal's point of view given the information available to her}. We denote this action by \(\tilde{a}_t\). In the first period, \(\tilde{a}_t\) is randomly drawn from \(\mathbf{A}_{P1}\), while for all further periods, it is decided upon in \textit{sub-model C}.
In every period, the principal computes the premium parameter \(\phi_t\) according to
\begin{equation}
\phi_t=\argmax_{\phi\in[0,1]} U_P(\tilde{x}_{Pt},s(\tilde{x}_{Pt}))~, 
\end{equation}
\noindent where the expected outcome is computed according to Eq. \ref{eq:exp-outcome-principal}, and the sharing rule follows Eq. \ref{eq:sharing-rule}. Together with the task to be carried out by the agent, the sharing rule (including the premium parameter \(\phi_t\)) is part of the contract.

\subsubsection{Sub-model B}
\label{sec:submodel-b}
This sub-model captures the agent's decision whether or not to accept the contract in \(\tau=2\) and which action to take in \(\tau=3\). The contract offered to him in \(\tau=1\) is the input to this sub-model. 
To make his decision, the agent computes the action that maximizes his utility in period \(t\) according to 
\begin{equation}
a_t=\argmax_{a^\prime\in \textcolor{black}{\mathbb{R}^{+}}} U_A(s(\tilde{x}_{At}, a^\prime))~.
\end{equation}
The outcome $\tilde{x}_{At}$ and the sharing rule $s(\cdot)$ follow Eqs. \ref{eq:expected-outcome-agent} and \ref{eq:sharing-rule}, respectively. If \(U_A(s(\tilde{x}_{At}, a_t)) \geq \underline{U}\), the agent accepts the contract in \(\tau=2\) and takes action \(a_t\) in \(\tau=3\).

\subsubsection{Sub-model C}
\label{sec:submodel-c}
Sub-model C includes the decision rules employed by the principal in \(\tau=6\) and \(7\) (see Fig. \ref{fig:abm}). The decision as to whether to perform a local or a global search is driven by the principal's search tendency \(\delta \in [0,1]\) and her estimation of the environment's effect in the previous period, \(\tilde{\vartheta}_{t-1}\). The \textcolor{black}{threshold for a global search} in \(t\), \(\kappa_t\), is implicitly defined by 
\begin{equation}
\delta = \dfrac{1}{\sigma \cdot \sqrt{2\pi}} \cdot \int_{-\infty}^{\kappa_t} e^{-\frac{1}{2}\cdot\left(\dfrac{z-\mu}{\sigma}  \right)} dz ~,
\label{eq:3}
\end{equation}
where the mean is $\mu = \mu(\mathbf{M}_{Pt})$, and the standard deviation is $\sigma=\sigma(\mathbf{M}_{Pt})$. The principal performs a global (local) search for alternative actions if \(\tilde{\vartheta}_{t-1} \geq \kappa_t\) (\(\tilde{\vartheta}_{t-1} < \kappa_t\)) \citep{Leitner2020}. \textcolor{black}{The rationale behind this modelling choice is that the principal is modelled to be more likely to search in the global search space when she is \enquote*{unsatisfied} with the outcome, i.e., when the environment has severe negative effects on the outcome. Whether or not she is unsatisfied -- and, in consequence, whether or not she searches globally -- is controlled via the parameter $\delta$, from which the threshold $\kappa_t$ in period $t$ is derived following Eq. \ref{eq:3}.}

In steps \(\tau=6\) and \(7\) of period $t$, the principal searches for the action \(\tilde{a}_{t+1}\) \textcolor{black}{that is utility maximizing given the information currently available to her} (see Fig. \ref{fig:abm}). The search spaces are defined relative to each other: The local search space captures a fraction of \(1/\lambda\) of \(\mathbf{A}_{Pt}\) and is equally distributed around \(\tilde{a}_{t}\) (the action \textcolor{black}{for which the principal constructed the incentives} in round \(t\)).\footnote{If \(\tilde{a}_t\) is located outside of \(\mathbf{A}_{Pt}\), the principal is forced to search globally for alternative actions.} The global search space is the fraction \(1-(1/\lambda)\) of \(\mathbf{A}_{Pt}\) and is the area outside of the space for a local search but inside of the boundaries of \(\mathbf{A}_{Pt}\). Whether the principal searches in the global or the local search space is driven by her search tendency, so that higher (lower) values of $\delta$ increase (decrease) the probability of a global search.
Once the search space is fixed, the principal randomly finds two alternative actions (with uniformly distributed probability) that are evaluated together with \(\tilde{a}_t\). She determines the action that promises to maximize her utility (see Eq. \ref{eq:principal}) as the action \(\tilde{a}_{t+1}\), which is the input into \textit{sub-model A} in round \(t+1\).

\begin{table}
\centering
\begin{tabular}{>{\centering\arraybackslash}p{0.12\textwidth}p{0.5\textwidth} p{0.3\textwidth}}
Symbol		&	Description		& Range/default value\\
\hline\noalign{\smallskip} 
\multicolumn{3}{l}{\textbf{Parameters related to the principal}} \\
\(m\)			& 		Access to previous estimations of the environment	&	\(\{1, 3, \infty\}\)			\\
\(1/\lambda\)	&		Fraction of the action space that can be accessed in case of a local search	&	\(\{1/3, 1/5, 1/10\}\)    \\
\(\delta\)		&		Tendency for global search								&	\(\{.25, .50, .75\}\)	\\	
\multicolumn{3}{l}{\textbf{Parameters related to the agent}} \\
\(m	\)		& 		Access to previous observations of the environment	&	\(\{1, 3, \infty\}\)			\\
\(\rho\)		&		Productivity											&	\(50\)		\\
\(\eta\)		&		Arrow-Pratt measure of risk-aversion							&	\(.50\)	\\
\multicolumn{3}{l}{\textbf{Parameters related to the environment}} \\
\(\sigma\)		&		Standard deviation, defined relative to the optimal performance \(x^{\ast}\) suggested by the static hidden-action model									&	\(\{.05x^{\ast}, .25x^{\ast}, \newline .45x^{\ast} ,.65x^{\ast} \}\)\\
\(\mu\)		&		Mean												& 	\(0\)		\\
\multicolumn{3}{l}{\textbf{Global parameters}} \\	
\(T\)			&		Periods							&	\(20\)	\\
\(R\)			&		Simulation runs										&	\(700\) \\
\hline\noalign{\smallskip} 
\end{tabular}
\caption{Key parameters of the agent-based model and ranges for simulation experiments}
\label{tab:parameters}
\end{table}

\section{Results}
\label{sec:results}

\subsection{Simulation experiments}

We perform simulation experiments for principals and agents who are characterized by having either a low, medium, or high memory, by setting the memory parameter $m$ to a value of $1$, $3$, and $\infty$, respectively. For the principal, we consider the cases of access to small ($1/\lambda = 1/10$), medium ($1/\lambda = 1/5$), and large local search spaces ($1/\lambda = 1/3$) and three different search tendencies. For the latter, we include a strong tendency for \textcolor{black}{a local search} ($\delta=0.25$), indifferent principals who perform \textcolor{black}{a local} and \textcolor{black}{global search} with the same probability ($\delta=0.50$), and principals who are more prone to \textcolor{black}{a global search} ($\delta=0.75$) into our analysis. In addition, we consider environments of different \textcolor{black}{turbulence}. In particular, we control for \textcolor{black}{environmental turbulence} via the standard deviation of the distribution of the exogenous variable that we set relative to the optimal outcome \textcolor{black}{$x^{\ast} = a^{\ast} \cdot \rho + \mu$ (see Eq. \ref{eq:prod-function}, the expected value of the environmental variable is equal to the mean of the environmental variables' distribution $\mu$).} \textcolor{black}{The optimal effort level is indicated by $a^{\ast}$ and defined by
\begin{subequations}
\begin{align}
a^{\ast} \in \argmax_{a\in \mathbb{R}^{+},\phi \in [0,1]} 	\quad 	& 	{E}\left(U_P\left(x-s\left(x\right)\right)\right) \label{eq:opta}\\
\textrm{s.t.} 	\quad 	&	{E}\left(U_A\left(s\left(x\right),a\right)\right)\geq\underline{U} \label{eq:opta:PC}\\
 					 &	a \in \argmax_{a^\prime \in \mathbb{R}^{+}}  {E} \left(U_A\left(s\left(x\right),a^\prime\right)\right)~, \label{eq:opta:ICC}
\end{align}
\end{subequations}
\noindent whereby the principal's and the agent's utility functions, the production function, and the sharing rule are the ones used in the agent-based model variant (see Tab. \ref{tab:comparison}). For the sake of readability and since the optimal effort $a^{\ast}$ is the same for every period $t$, we suppress the notion of $t$ in Eqs. \ref{eq:opta} to \ref{eq:opta:ICC}.}
We set the standard deviation at either $\sigma=0.05x^{\ast}$,  $0.25x^{\ast}$, $0.45x^{\ast}$ or $0.65x^{\ast}$, indicating a range from relatively stable to relatively turbulent environments. These parameter settings result in a total number of $3 \cdot 3 \cdot 3 \cdot 4 =108$ scenarios. We perform $R=700$ simulation runs for every scenario, and our analysis focuses on the first $T=20$ periods.\footnote{The number of simulation runs was fixed on the basis of variance analysis \citep{lorscheid2012opening}.} Table \ref{tab:parameters} summarizes the key parameters included in the model and the global simulation parameters.

\subsection{Dynamics in the principal's domain}
\label{sec:macrodynamics}

\subsubsection{Dynamics in the principal's search for optimal incentive schemes: The Sisyphus effect} 
\label{sec:macroscopic-effect}

\textcolor{black}{We refer to the dynamics in the principal's sphere as the Sisyphus effect. Recall that the principal searches for the optimal sharing rule with limited \textcolor{black}{access to} information about the action space and the environment. Over time, the sharing rule offered to the agent becomes more and more similar to the optimal sharing rule, i.e., the rule derived under the assumption of unlimited \textcolor{black}{access to} information. Due to limited \textcolor{black}{access to} information, the principal might overshoot the target of the optimal sharing rule. In our analogy, the principal's search for the optimal sharing rule corresponds to the Greek prisoner Sisyphus rolling a heavy boulder up a mountain. However, due to limited \textcolor{black}{access to} information, the principal cannot perfectly locate the top of the mountain, and she might keep rolling the boulder even though she has already reached the top. In consequence, she sees the boulder rolling back down the mountain, only to start over the search procedure again. Please note that our concept of the Sisyphus effect slightly differs from the Sisyphus effect defined for the domain of consumer behavior, which is that decision-makers who are maximizers tend to minimize the importance of previous decisions \citep{Simon1957}. In consequence, decision-makers do not include past experiences in their search; for every decision -- even if it is highly related to previously made decisions -- they always start their search at the bottom of the mountain \citep{schwartz2002}. Thus, while the concept of the Sisyphus effect in the domain of consumer behavior describes decision makers who do not capitalize on previous knowledge acquired and therefore start over their search with every decision, our concept of the Sisyphus effect leads to starting over the search procedure due to imprecise information about the optimal choice, i.e., about the top of the mountain.}

\begin{figure}
\center
  \includegraphics[width=1\linewidth]{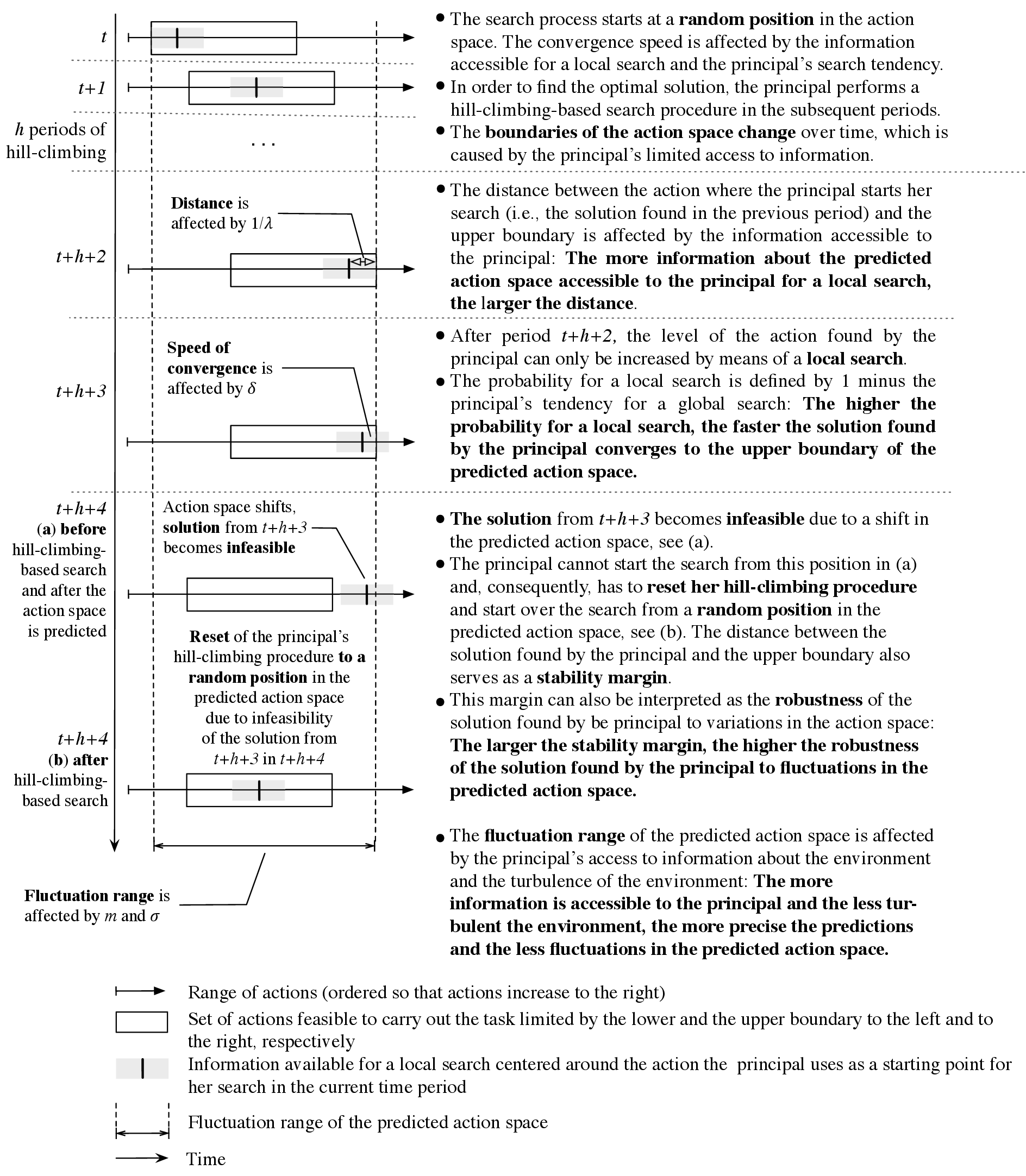}
  \caption{Schematic representation of the dynamics leading to the Sisyphus effect}
  \label{fig:macroeffect}
\end{figure}

The dynamics of \textcolor{black}{this effect} are driven by two forces:
\begin{enumerate*}[label=\textit{(\roman*)}]
\item The distance between the solution found by the principal and the optimal one, and
\item the fluctuation range of the predicted action space.
\end{enumerate*} The dynamics unfolding from the interaction of these two forces are illustrated in Fig. \ref{fig:macroeffect}, which schematically shows a stylized hill-climbing procedure with a global search: In period \(t\), the principal starts at a random position in the predicted action space (indicated by the vertical line). The gray area around this position indicates the local search space. In period \(t+1\), the principal performs a global search outside of the gray area but inside the boundaries of the action space, leading to an increase in the \textcolor{black}{agent's effort}. The search procedure is repeated \(h\) times until the discovered solution is close to the action space's upper boundary.\footnote{Recall that actions are ordered, and due to first-order stochastic dominance, higher action levels lead to higher outcomes. Therefore, to increase the outcome, the principal intends to move towards the upper boundary of the action space.} 

The driving forces \textit{(i)} and \textit{(ii)} interact in two ways. First, the distance between the solution found by the principal and the optimal solution in period \(t+h+2\) is influenced by the size of the local search space (\(1/\lambda\)). The number of periods required to come up with close-to-optimal solutions is affected by \(1/\lambda\) and the principal's search tendency \(\delta\). Recall, the latter influences whether a local or a global search is performed. The principal follows a hill-climbing strategy and intends to increase the level of the action stepwise by selecting actions that \textcolor{black}{are} closer to the optimal action \citep{Cormen2009}. The search spaces are defined relative to each other: If the information accessible to the principal in a local search increases from \(1/10\) to \textcolor{black}{\(1/3\)} of the entire action space, the information accessible in case of a global search decreases from \(9/10\) to \(2/3\) of the entire action space, respectively. Thus, as the size of the \textit{local} search space increases, the principal is forced to be more innovative -- in terms of longer jumps -- when performing a \textit{global} search. 

A strong focus on long jumps (in terms of a high \(\delta\) and large \(1/\lambda\)) is efficient as long as the solution can be improved in significant steps \citep{Yang2007}, which is usually the case at the beginning of a search procedure. A focus on short jumps at the beginning of a search procedure -- in terms of low \(\delta\) and \(1/\lambda\) --, in contrast, will result in a slower \textcolor{black}{increase in the agent's effort}. In Fig. \ref{fig:macroeffect}, the periods required by the principal for searching for better solutions are indicated by \(h\). As soon as the principal comes up with solutions that are already close to the optimal one (see periods \(t+h+2\) in Fig. \ref{fig:macroeffect}), a strong emphasis on \textcolor{black}{a global search} will make it difficult for her to find better solutions since there is no room for improvement outside the local search space \citep{Yao1999}. Thus, an exclusive focus on a global search in these periods would result a standstill, which has a conceptual similarity to a local maximum.\footnote{Information about a small (large) fraction of the action space indicates that the principal comes to a standstill closer (farther) to the optimal solution.} A local search is the only way out of the \enquote*{local maximum} (see period \(t+h+3\) in Fig. \ref{fig:macroeffect}). Consequently, the  speed \textcolor{black}{at which the agent's effort increases} in periods \(t+h+2\) onward is driven by the principal's search tendency \(\delta\), so that a strong (weak) tendency for a local search leads to a faster (slower) convergence to the optimal solution.

Second, the fluctuation range of the predicted action space is affected by the principal's memory and the turbulence in the environment: Recall that the principal searches in the set of actions feasible to carry out the task when performing her hill-climbing-based search and that her prediction of the environment is included in the lower and the upper boundaries of the predicted action space (see Sec. \ref{sec:p-and-a-info-access}). The fluctuation of the action space is affected by the precision of the principal's predictions, whereby more precision directly translates to less fluctuation in the lower and the upper boundaries.
In the model introduced here, the stability of the principal's prediction is affected by the access to information about the environment (\(m\)) and the turbulence of the environment (\(\sigma\)), such that low values of \(m\) and/or high values of \(\sigma\) lead to relatively large fluctuation ranges. In contrast, the opposite holds true for rather large values of \(m\) and/or high values of \(\sigma\).

These interactions reveal some interesting dynamics: \textcolor{black}{The speed at which the agent's effort increases over time is affected by}  the size of the local search space and the principal's search tendency. 
For the principal, solutions close to the optimal one are beneficial as long as she can predict the action space precisely. Suppose the precision of the principal's prediction of the action space is low. In that case, the fluctuation range of the \textcolor{black}{predicted action} space is high, which increases the likelihood of the action from which the principal starts her search shifting into the infeasible region (see period \(t+h+4\) (a) in Fig. \ref{fig:macroeffect}). 
Please note that the principal is unaware of the boundaries of the action space but can only assess whether an action is feasible or not and cannot evaluate the distance to the feasible region.\footnote{\textcolor{black}{See the assumptions regarding information access in the agent-based model variant introduced in Tab. \ref{tab:info-access}.}} 
Therefore, she is forced to {reset her hill-climbing procedure}, i.e., she moves the action level to a random position in the predicted action space to start over the search process (see also period \(t+h+4\) (b) in Fig. \ref{fig:macroeffect}). 

From these dynamics, we can identify the following trade-off: \textcolor{black}{Moving towards} the optimal solution faster is desirable for the principal as it is associated with a higher \textcolor{black}{effort level}. This, however, comes at the cost of robustness. Suppose the distance between the found solution and the optimal one is small (large). In that case, the solution is less (more) robust to variations in the upper boundary, and it is more (less) likely that the found solution becomes infeasible if the upper boundary shifts to the left (see Fig. \ref{fig:macroeffect}). We refer to the distance between the solution bound by the principal and the upper boundary of the action space as the \textit{stability margin}. 

\subsubsection{Limited information and final performances} 
\label{sec:final-performance}

\paragraph{Performance measure} To analyze how well the incentive scheme effective by the end of the observation period performs, we report the mean normalized actions taken by the agent in the last period as a performance measure in Tab. \ref{tab:results-final} (together with the confidence intervals at the \(99\%\)-level).\footnote{The control of the agent's behavior is at the center of interest in hidden-action setups. The action is directly related to the outcome via the production function. Compared to the outcome, analyzing the action taken by the agent is free from any environmental impact. Therefore, basing the analysis on the actions taken by the agent appears appropriate \citep[see also][]{Leitner2020}.} For period \(t=20\), the mean normalized action taken by the agent is computed according to  \begin{equation}
\tilde{a}_{t}=\frac{1}{R} \sum_{r=1}^{r=R} \frac{a_{tr}}{a^{\ast}}~,
\label{eq:macro-effects}
\end{equation}
where \(a^{\ast}\) represents the optimal action \textcolor{black}{(see Eqs. \ref{eq:opta} to \ref{eq:opta:ICC})}, and \(a_{tr}\) stands for the action taken by the agent in period \(t\) and simulation round \(r\).

\begin{sidewaystable}
\begin{scriptsize}
\flushleft
\begin{tabular}{l l l l l l l l l l l l l l l}
&&&\multicolumn{12}{c}{Environmental turbulence (\(\sigma\))} \\
\noalign{\smallskip} \cline{4-15} \noalign{\smallskip}\noalign{\smallskip}
&&&\multicolumn{3}{c}{\(.05x^{\ast}\)} & \multicolumn{3}{c}{\(.25x^{\ast}\)} & \multicolumn{3}{c}{\(.45x^{\ast}\)} & \multicolumn{3}{c}{\(.65x^{\ast}\)}  	\\	
\noalign{\smallskip} \cline{4-15} \noalign{\smallskip}\noalign{\smallskip}
\multirow{2}{*}{\shortstack[l]{Principal's\\tendency for a\\global search (\(\delta\))} }	&
\multirow{2}{*}{\shortstack[l]{Information\\about envi-\\ronment (\(m\))}}	 & &
\multicolumn{3}{c}{\shortstack{Information about\\action space (\(1/\lambda\))}} & \multicolumn{3}{c}{\shortstack{Information about\\action space (\(1/\lambda\))}} & \multicolumn{3}{c}{\shortstack{Information about\\action space (\(1/\lambda\))}}  & \multicolumn{3}{c}{\shortstack{Information about\\action space (\(1/\lambda\))}}   \\
		&&& \(1/10\)	& \(1/5\)		& \(1/3\) 	& \(1/10\)	& \(1/5\)		& \(1/3\) & \(1/10\)	& \(1/5\)		& \(1/3\) & \(1/10\)	& \(1/5\)		& \(1/3\) \\
		\noalign{\smallskip} \hline \noalign{\smallskip}		
		 \multirow[t]{6}{*}{{.25}} 
		&\multirow[t]{2}{*}{{1}} 
		& \(\tilde{a}_{20}\) 		& \(.737\) 	& \(.824\) & \(.819\) & \(.681\) & \(.692\) & \(.709\) & \(.652\) & \(.673\) & \(.668\) & \(.678\) & \(.715\) & \(.696\) \\
		& & CI	& \(\pm.017\) 	& \(\pm.016\) & \(\pm.017\) & \(\pm.025\) & \(\pm.025\) & \(\pm.025\) & \(\pm.032\) & \(\pm.032\) & \(\pm.030\) & \(\pm.037\) & \(\pm.038\) & \(\pm.039\) \\
		&\multirow[t]{2}{*}{{3}} 
		& \(\tilde{a}_{20}\) 			 & \(.892\) 	& \(.891\) & \(.901\) & \(.823\) & \(.839\) & \(.832\) & \(.789\) & \(.789\) & \(.792\) & \(.767\) & \(.767\) & \(.795\) \\
		& & CI		 &\(\pm.013\) & \(\pm.013\) & \(\pm.014\) & \(\pm.019\) & \(\pm.018\) & \(\pm.019\) & \(\pm.023\) & \(\pm.023\) & \(\pm.024\) & \(\pm.025\) & \(\pm.026\) & \(\pm.027\) \\
		&\multirow[t]{2}{*}{{\(\infty\)}} 
		& \(\tilde{a}_{20}\) 			 & \(.945\) 	& \(.954\) & \(.962\) & \(.898\) & \(.915\) & \(.917\) & \(.901\) & \(.893\) & \(.902\) & \(.877\) & \(.885\) & \(.893\) \\
		& & CI		 &  \(\pm.011\) & \(\pm.010\) & \(\pm.009\) & \(\pm.013\) & \(\pm.012\) & \(\pm.013\) & \(\pm.014\) & \(\pm.015\) & \(\pm.014\) & \(\pm.015\) & \(\pm.015\) & \(\pm.015\) \\
		\noalign{\smallskip} \hline \noalign{\smallskip}
		\multirow[t]{6}{*}{{.50}} 
		&\multirow[t]{2}{*}{{1}} 
		& \(\tilde{a}_{20}\)  			 &  \(.743\) & \(.824\) & \(.829\) & \(.650\) & \(.707\) & \(.706\) & \(.667\) & \(.678\) & \(.693\) & \(.677\) & \(.711\) & \(.696\) \\
		& & CI		 &  \(\pm.017\) & \(\pm.016\) & \(\pm.016\) & \(\pm.026\) & \(\pm.025\) & \(\pm.025\) & \(\pm.029\) & \(\pm.031\) & \(\pm.032\) & \(\pm.038\) & \(\pm.040\) & \(\pm.039\) \\
		&\multirow[t]{2}{*}{{3}} 
		& \(\tilde{a}_{20}\)  			 & \(.901\) & \(.909\) & \(.908\) & \(.843\) & \(.839\) & \(.835\) & \(.816\) & \(.810\) & \(.822\) & \(.789\) & \(.812\) & .789 \\
		& & CI		 & \(\pm.012\) & \(\pm.011\) & \(\pm.012\) & \(\pm.017\) & \(\pm.018\) & \(\pm.019\) & \(\pm.023\) & \(\pm.022\) & \(\pm.023\) & \(\pm.025\) & \(\pm.027\) & \(\pm.027\) \\
		&\multirow[t]{2}{*}{{\(\infty\)}} 
		& \(\tilde{a}_{20}\) 		 & \(.960\) & \(.964\) & \(.966\) & \(.925\) & \(.925\) & \(.929\) & \(.899\) & \(.918\) & \(.903\) & \(.899\) & \(.902\) & \(.908\) \\
		& & CI		 & \(\pm.008\) & \(\pm.008\) & \(\pm.007\) & \(\pm.011\) & \(\pm.011\) & \(\pm.010\) & \(\pm.014\) & \(\pm.011\) & \(\pm.014\) & \(\pm.015\) & \(\pm.014\) & \(\pm.013\) \\
		\noalign{\smallskip} \hline \noalign{\smallskip}
		\multirow[t]{6}{*}{{.75}} 
		&\multirow[t]{2}{*}{{1}} 
		& \(\tilde{a}_{20}\) 			 & \(.733\) & \(.828\) & \(.843\) & \(.659\) & \(.700\) & \(.707\) & \(.653\) & \(.669\) & \(.688\) & \(.683\) & \(.692\) & \(.710\) \\
		& & CI		 & \(\pm.019\) & \(\pm.016\) & \(\pm.016\) & \(\pm.025\) & \(\pm.025\) & \(\pm.025\) & \(\pm.030\) & \(\pm.032\) & \(\pm.031\) & \(\pm.038\) & \(\pm.037\) & \(\pm.038\) \\
		&\multirow[t]{2}{*}{{3}} 
		& \(\tilde{a}_{20}\) 			 &\(.917\) & \(.924\) & \(.916\) & \(.863\) & \(.851\) & \(.833\) & \(.817\) & \(.808\) & \(.825\) & \(.802\) & \(.794\) & \(.804\) \\
		& & CI		 &  \(\pm.010\) & \(\pm.009\) & \(\pm.010\) & \(\pm.015\) & \(\pm.017\) & \(\pm.017\) & \(\pm.023\) & \(\pm.022\) & \(\pm.022\) & \(\pm.026\) & \(\pm.026\) & \(\pm.026\) \\
		&\multirow[t]{2}{*}{{\(\infty\)}} 
		& \(\tilde{a}_{20}\) 			 & \(.961\) & \(.967\) & \(.956\) & \(.932\) & \(.935\) & \(.925\) & \(.915\) & \(.915\) & \(.909\) & \(.912\) & \(.904\) & \(.899\) \\
		& & CI		 & \(\pm.007\) & \(\pm.005\) & \(\pm.006\) & \(\pm.009\) & \(\pm.009\) & \(\pm.009\) & \(\pm.011\) & \(\pm.011\) & \(\pm.011\) & \(\pm.012\) & \(\pm.013\) & \(\pm.012\) \\
\hline\noalign{\smallskip}
\end{tabular}\\
\(\tilde{a}_{20}\)=mean normalized action taken by the agent in period \(t=20\) (see Eq. \ref{eq:macro-effects}), CI=confidence interval for \(\alpha=.01\).
\caption{Mean performance and confidence intervals after 20 periods}
\label{tab:results-final}
\end{scriptsize}
\end{sidewaystable}

\paragraph{Results} 
The Sisyphus effect introduced in Sec. \ref{sec:macroscopic-effect} substantially shapes the results presented in Tab. \ref{tab:results-final}. We identify three main results.
First, the optimal incentive scheme does \textit{not} emerge if access to information is limited.  As discussed in Sec. \ref{sec:macroscopic-effect}, even if the principal can find a solution almost identical to the optimal one, the slightest fluctuation in the action space's predicted upper boundary leads to a reset of the hill-climbing procedure. Thus, it is rather unlikely that close-to-optimal solutions are perpetuated for a long time.  

Second, we can observe that an increase in access to information about the environment results in higher action levels in all scenarios.  
This effect appears to be more pronounced if the local search space is small. This is intuitive, as more information about the environment leads to rather precise predictions of the action space, and small local search spaces result in relatively close solutions to the optimal one. In rather turbulent environments, this finding is less pronounced (but still significant). Since turbulence in the environment decreases the precision of the predicted action space's upper boundary, the Sisyphus effect becomes more severe in these cases, which results in lower action levels \textit{on average}.

Third, it is in line with the Sisyphus effect that large local search spaces are particularly beneficial for the final performance if the principal puts a strong focus on \textcolor{black}{\textit{a local search}} (\(\delta=.25\)). 
However, it is counter-intuitive that large local search spaces are even more beneficial if the principal puts a strong emphasis on \textcolor{black}{\textit{a global search} instead of a local search} (\(\delta=.75\)). 
Two competing effects drive this last observation: First, there are performance drops caused by the Sisyphus effect, and second, moderately well-performing solutions can be perpetuated for a longer period of time, since larger local search spaces translate into larger stability margins (Sec. \ref{sec:macroscopic-effect}), which is why the organization is better off \textit{on average}. In relatively turbulent environments, the former effect overrules the latter one, while the opposite holds for relatively stable environments.\footnote{In this context, the principal's search tendency appears to have a slightly moderating effect. We already know that the search tendency drives the speed at which the solution converges to the upper boundary of the action space (after period \(t+h+2\) in Fig. \ref{fig:macroeffect}). 
Suppose the probability of a local search is low (high). In that case, the convergence of the found solution to the upper boundary of the action space is decelerated (accelerated), making it less (more) likely for the Sisyphus effect to come into force.} 

\subsubsection{Limited information and patterns in performance over time}
\label{sec:macro-patterns}

\begin{figure}
\center
  \includegraphics[width=\linewidth]{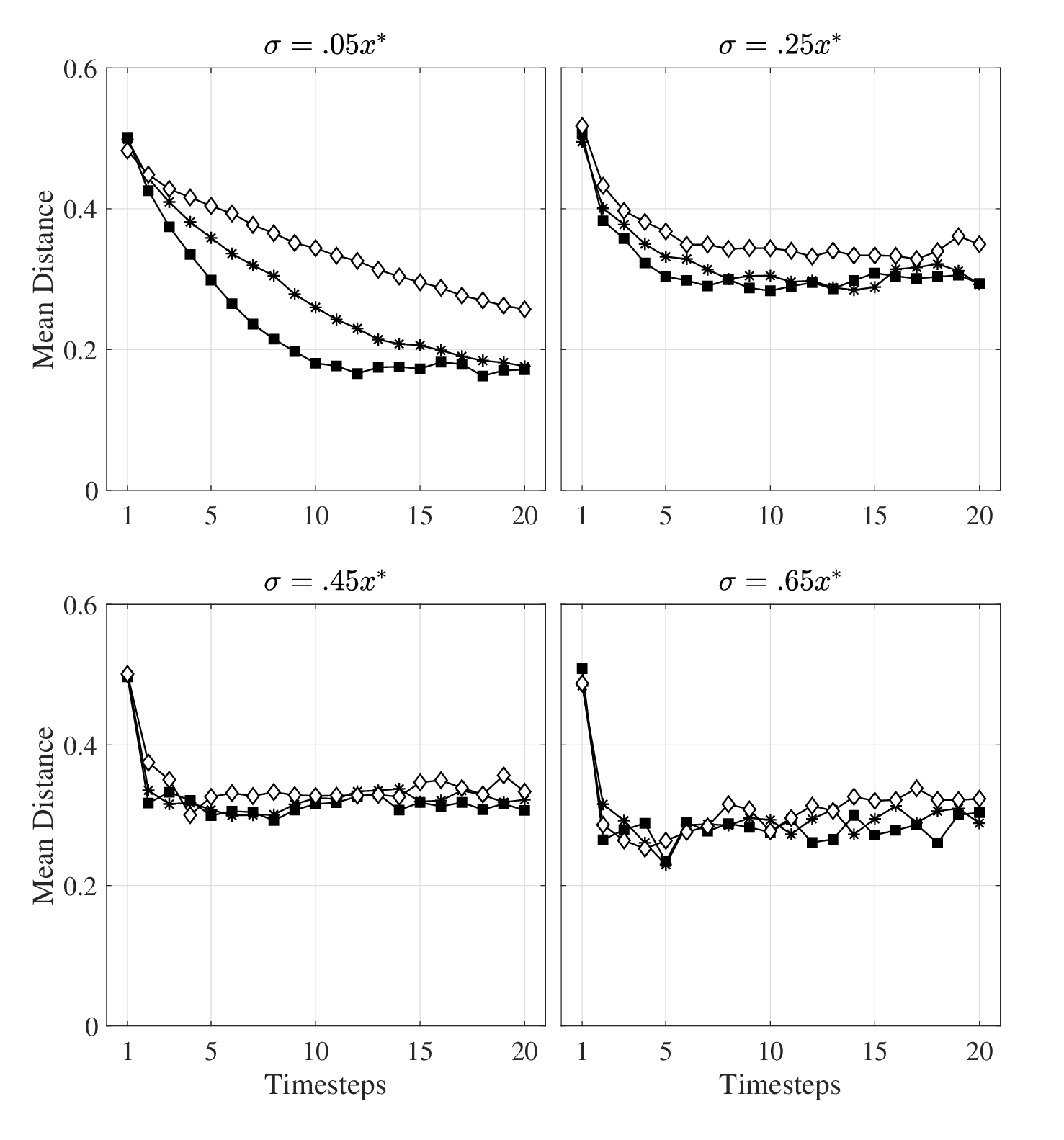}
  \caption{\textcolor{black}{Mean} distances emerging from variations in access to information for a local/global search for different degrees of environmental turbulence. The principal's tendency for a global search is set \(\delta =.5\) and access to information about the environment is set to \(m=1\). Variations in access to information for a local/global search are represented by black squares (\(\blacksquare\)) for \(1/\lambda = 1/3\), asterisks (\(\Asterisk\)) for \(1/\lambda = 1/5\), and white diamonds (\(\lozenge\)) for \(1/\lambda = 1/10\).}
  \label{fig:actionspace_access}
\end{figure}

\begin{figure}
\center
  \includegraphics[width=\linewidth]{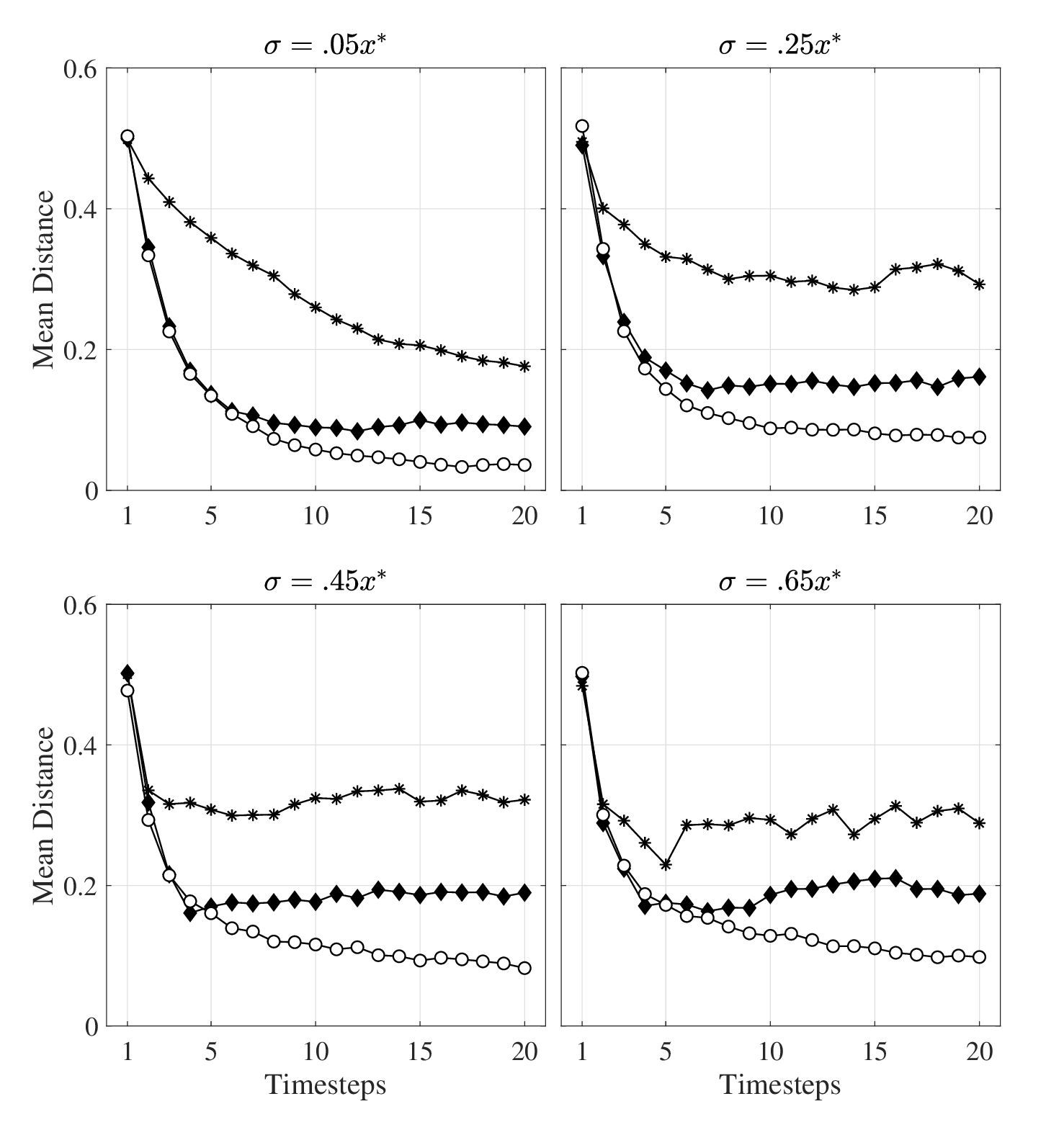}
    \caption{\textcolor{black}{Mean} distances emerging from variations in access to information about the environment for different degrees of environmental turbulence. Plotted lines represent the sum of the squared distances (see Eq. \ref{eq:euclidean-distance}) for scenarios in which the principal's tendency for a global search is set \(\delta =.5\) and access to information for local/global search is set to \(1/\lambda=1/5\). Variations in access to information about the environment are represented by asterisks (\(\Asterisk\)) for \(m=1\), black diamonds (\(\blacklozenge\)) for \(m=3\), and white circles (\(\circ\)) for \(m=\infty\).}
  \label{fig:environment_access}
\end{figure}

\paragraph{Performance measure}  We complement the analysis provided in Sec. \ref{sec:final-performance} by studying the patterns emerging from the principal's search under limited access to information over time. To do so, we plot the \textcolor{black}{mean distance} \(d_t\) between the mean normalized action (Eq. \ref{eq:macro-effects}) and the optimal action under unlimited \textcolor{black}{access to} information \textcolor{black}{(Eqs. \ref{eq:opta} to \ref{eq:opta:ICC})} in Figs. \ref{fig:actionspace_access} and \ref{fig:environment_access}. The distance measure \(d_t\) is computed according to Eq. \ref{eq:euclidean-distance}:\

\begin{equation}
d_t = \frac{1}{R}\sum_{r=1}^{r=R} (1-\frac{a_{tr}}{a^{\ast}})~.
\label{eq:euclidean-distance}
\end{equation}

\paragraph{Results} The analysis includes cases with \(\delta=0.50\), which indicates that the principal is indifferent as to whether to perform a local or a global search.\footnote{The patterns emerging for scenarios with \(\delta=0.25\) and \(0.75\) are almost identical to the pattern presented here, which is why we present the results for \(\delta=0.50\) only.} 
Figure \ref{fig:actionspace_access} presents the patterns emerging from the variations in the size of the local search space (parameter \(1/\lambda\)) for four different levels of environmental turbulence, \(\sigma=\{0.05x^{\ast}, 0.25x^{\ast}, 0.45x^{\ast}, 0.65x^{\ast}\}\). For the results presented in Fig. \ref{fig:actionspace_access}, the principal's access to information about the environment is fixed to \(m=1\) (for further variation in \(m\), see Fig. \ref{fig:environment_access}).

For mid-stable and turbulent environments \(\sigma=\{0.25x^{\ast}, 0.45x^{\ast}, 0.65x^{\ast}\}\), almost identical patterns can be observed, while for stable environments \(\sigma=0.05x^{\ast}\)
the results indicate that larger local search spaces lead to a faster speed at which performance increases. 
In cases in which a fraction of \(1/\lambda=1/3\) is accessible (indicated by black squares), a distance measure close to \textcolor{black}{$0.2$} can be observed after eight periods, while for cases in which \(1/\lambda=1/5\) of the action space can be accessed for a local search (indicated by asterisks), the same distance can be achieved after only around \textcolor{black}{16} periods. For the case of \(1/\lambda=1/10\) (indicated by white diamonds), we can observe that the mean normalized action \textcolor{black}{moves towards} the optimal one very slowly. Thus, in stable environments, the larger stability margins caused by larger local search spaces affect the final performance (see Sec. \ref{sec:final-performance}) and the speed \textcolor{black}{at which the agent's action moves towards the optimal one}. However, turbulent environments and, as a consequence, more fluctuations in the predicted action space offset the positive effects of larger stability margins.

Figure \ref{fig:environment_access} presents the patterns of the distance measure emerging over time for variations in access to information about the environment (parameter \(m\)). The parameter controlling the principal's search tendency is again set to \(\delta=.50\), and the access to information for a local search is set to the medium level of \(1/\lambda=1/5\). For the scenarios with medium to high turbulence in the environment, i.e., \(\sigma=\{0.25x^{\ast}, 0.45x^{\ast}, 0.65x^{\ast}\}\), we can observe that the distance measures almost stabilize after only around five periods at the levels of final performance discussed in Sec. \ref{sec:final-performance}.
 
For stable environments (\(\sigma=0.05x^{\ast}\)), we can observe that for unlimited access to environmental information (\(m=\infty\), indicated by circles), the distance measures are close to zero. This is an intuitive finding because due to unlimited \textcolor{black}{access to} information and the stable environment the predicted action space only marginally fluctuates, and the principal's hill-climbing-based search brings her very close to the optimal solution. As a consequence, the Sisyphus effect is less severe. For minimal memory (\(m=1\)), the results indicate that the severity of the Sisyphus effect increases substantially; consequently, the \textcolor{black}{agent's effort increases} significantly more slowly.

\subsection{Dynamics in the agent's domain}
\label{sec:microdynamics}

\subsubsection{Dynamics of the agent's actions: Excess effort} 
\label{sec:microscopic-effect}

In Sec. \ref{sec:macroscopic-effect}, it has been discussed that limited access to estimations of the environment may lead to inaccurate predictions of the future state of the environment by the principal. The same applies to the agent: Contrary to the principal, he can \textit{observe} the environment (see also Tab. \ref{tab:info-access}). However, since the access to previous observations might be limited, the agent's predictions of the environment might be imprecise, too. Recall, the agent includes his predictions in the computation of feasible actions to perform the task assigned to him. 

\begin{figure}
\center
  \includegraphics[width=1\linewidth]{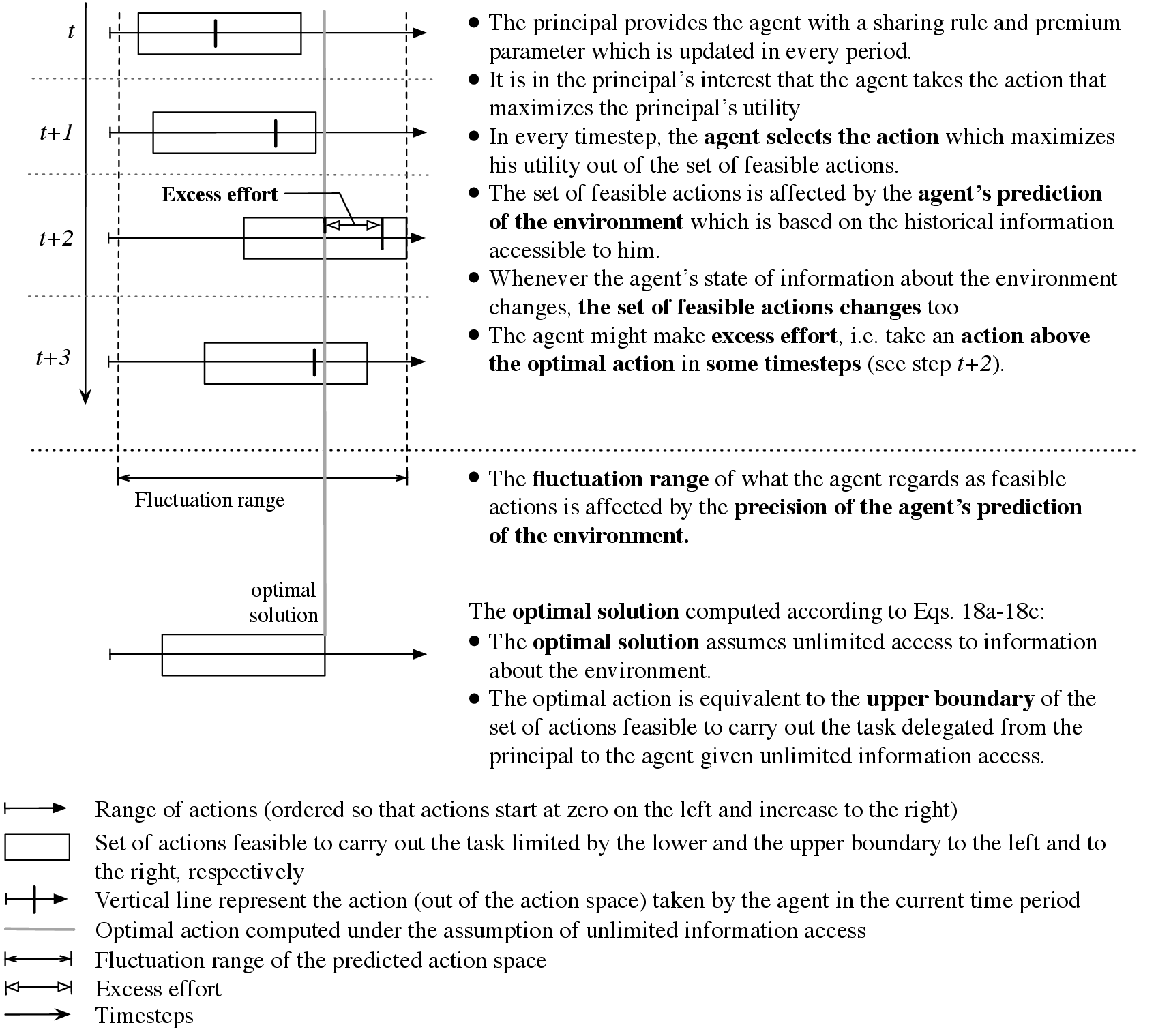}
  \caption{Schematic representation of the dynamics in the agent’s decisions for actions leading to excess effort}
  \label{fig:microeffect}
\end{figure}

A schematic representation of the agent's decisions is presented in Fig. \ref{fig:microeffect}. When predicting the environment, due to limited \textcolor{black}{access to} information and/or turbulence, the agent might over- or  underestimate the \textit{correct value} of the exogenous variable, which is why the predicted upper boundary of \textcolor{black}{the set of feasible actions} might be either above or below the \textit{correct} upper boundary (see Fig. \ref{fig:microeffect}). Recall that the hill-climbing-based search procedure employed by the principal and the computation of the premium parameter provides the agent with incentives to take actions that are close to what is perceived as the upper boundary. Consequently, if the predicted upper boundary of the action space is \textit{above} the optimal action, the agent likely makes even \textit{more} of an effort than optimal in the case of full information. We refer to this extra effort as \textit{excess effort}, whereby whether and how much excess effort is made is affected by the fluctuation range of the predicted action space. 

\subsubsection{Probability of and extent of excess effort}
\label{sec:micro-patterns}

\paragraph{Performance measures} We report two measures: First, the agent's probability of making an excess effort and, second, the average extent of excess effort made by the agent during the entire observation period. We formalize the probability of making an excess effort by 
\begin{equation}
\mathbb{P}({a_{x}}) = \mathbb{P}(a_{tr}>a^{\ast}) = \frac{\sum_{t=2}^{T} \sum_{r=1}^{R} [a_{tr}>a^{\ast}]}{R \cdot (T-1)}~. 
\label{eq:micro1}
\end{equation}
The numerator follows the Iverson bracket notation so that the logical proposition $P$ contained in the brackets is either \([P]=1\), if the proposition is satisfied, or \([P]=0\), otherwise \citep{Graham1994}. The agent's average excess effort is computed according to 
\begin{equation}
\tilde{a}_x = \frac{ \sum_{t=2}^{T} \sum_{r=1}^{R} (\tilde{a}_{tr}-1) \cdot [a_{tr}>a^{\ast}] }  {\sum_{t=2}^{T} \sum_{r=1}^{R} [a_{tr}>a^{\ast}]}~,
\label{eq:micro2}
\end{equation} 
where \(\tilde{a}_{tr}=a_{tr}/a^{\ast}\) represents the action taken by the agent in period \(t\) and simulation run \(r\) normalized by the optimal action \textcolor{black}{(Eqs. \ref{eq:opta} to \ref{eq:opta:ICC})}. Please note that we exclude the initial period from the computation of the indicators introduced in Eqs. \ref{eq:micro1} and \ref{eq:micro2}, as the action taken by the agent cannot exceed the optimal action in \(t=1\).\footnote{In \(t=1\), the agent is randomly assigned an initial action in the action space.}
\begin{sidewaystable}
\begin{scriptsize}
\flushleft

\begin{tabular}{l l l l l l l l l l l l l l l}
&&&\multicolumn{12}{c}{Environmental turbulence (\(\sigma\))} \\
\noalign{\smallskip} \cline{4-15} \noalign{\smallskip}\noalign{\smallskip}
&&&\multicolumn{3}{c}{\(.05x^{\ast}\)} & \multicolumn{3}{c}{\(.25x^{\ast}\)} & \multicolumn{3}{c}{\(.45x^{\ast}\)} & \multicolumn{3}{c}{\(.65x^{\ast}\)}  	\\	
\noalign{\smallskip} \cline{4-15} \noalign{\smallskip}\noalign{\smallskip}
\multirow{2}{*}{\shortstack[l]{Principal's\\tendency for a\\global search (\(\delta\))} }	&
\multirow{2}{*}{\shortstack[l]{Information\\about envi-\\ronment (\(m\))}}	 & &
\multicolumn{3}{c}{\shortstack{Information about\\action space (\(1/\lambda\))}} & \multicolumn{3}{c}{\shortstack{Information about\\action space (\(1/\lambda\))}} & \multicolumn{3}{c}{\shortstack{Information about\\action space (\(1/\lambda\))}}  & \multicolumn{3}{c}{\shortstack{Information about\\action space (\(1/\lambda\))}}   \\
		&&& \(1/10\)	& \(1/5\)		& \(1/3\) 	& \(1/10\)	& \(1/5\)		& \(1/3\) & \(1/10\)	& \(1/5\)		& \(1/3\) & \(1/10\)	& \(1/5\)		& \(1/3\) \\
		\noalign{\smallskip} \hline \noalign{\smallskip}		
		 \multirow[t]{6}{*}{{.25}} 
		&\multirow[t]{2}{*}{{1}} 
		& \(\mathbb{P}(a_x)\)			& \(.012\) 	& \(.034\) & \(.045\) & \(.037\) & \(.050\) & \(.076\) & \(.122\) & \(.119\) & \(.138\) & \(.193\) & \(.194\) & \(.212\) \\
		& & \(\tilde{a}_{x}\)		& \(.013\) 	& \(.015\) & \(.013\) & \(.094\) & \(.073\) & \(.074\) & \(.211\) & \(.200\) & \(.190\) & \(.297\) & \(.316\) & \(.293\) \\
		&\multirow[t]{2}{*}{{3}} 
		& \(\mathbb{P}(a_x)\)				& \(.047\) 	& \(.053\) & \(.053\) & \(.082\) & \(.112\) & \(.129\) & \(.136\) & \(.151\) & \(.162\) & \(.163\) & \(.184\) & \(.203\) \\
		& & \(\tilde{a}_{x}\)		& \(.007\) 	& \(.009\) & \(.008\) & \(.040\) & \(.045\) & \(.044\) & \(.081\) & \(.083\) & \(.085\) & \(.130\) & \(.129\) & \(.137\) \\
		&\multirow[t]{2}{*}{{\(\infty\)}} 
		& \(\mathbb{P}(a_x)\)			 	& \(.077\) 	& \(.077\) & \(.052\) & \(.106\) & \(.140\) & \(.143\) & \(.157\) & \(.167\) & \(.175\) & \(.188\) & \(.195\) & \(.209\) \\
		& & \(\tilde{a}_{x}\)		& \(.005\) 	& \(.005\) & \(.005\) & \(.027\) & \(.025\) & \(.026\) & \(.047\) & \(.049\) & \(.052\) & \(.072\) & \(.075\) & \(.072\) \\
		\noalign{\smallskip} \hline \noalign{\smallskip}
		\multirow[t]{6}{*}{{.50}} 
		&\multirow[t]{2}{*}{{1}} 
		& \(\mathbb{P}(a_x)\)				& \(.018\) 	& \(.030\) & \(.047\) & \(.033\) & \(.050\) & \(.076\) & \(.120\) & \(.119\) & \(.130\) & \(.196\) & \(.206\) & \(.211\) \\
		& & \(\tilde{a}_{x}\)		&  \(.010\) 	& \(.014\) & \(.015\) & \(.100\) & \(.078\) & \(.071\) & \(.216\) & \(.208\) & \(.187\) & \(.299\) & \(.304\) & \(.299\) \\
		&\multirow[t]{2}{*}{{3}} 
		& \(\mathbb{P}(a_x)\)				&  \(.043\) 	& \(.048\) & \(.042\) & \(.107\) & \(.112\) & \(.125\) & \(.149\) & \(.160\) & \(.166\) & \(.202\) & \(.215\) &  \(.217\) \\
		& & \(\tilde{a}_{x}\)		& \(.008\) 	& \(.008\) & \(.007\) & \(.043\) & \(.045\) & \(.042\) & \(.085\) & \(.088\) & \(.090\) & \(.130\) & \(.135\) &  \(.132\) \\
		&\multirow[t]{2}{*}{{\(\infty\)}} 
		& \(\mathbb{P}(a_x)\)				&  \(.059\) 	& \(.050\) & \(.044\) & \(.131\) & \(.134\) & \(.121\) & \(.168\) & \(.184\) & \(.170\) & \(.204\) & \(.194\) & \(.200\) \\
		& & \(\tilde{a}_{x}\)		&  \(.005\) 	& \(.005\) & \(.005\) & \(.028\) & \(.028\) & \(.025\) & \(.047\) & \(.050\) & \(.053\) & \(.075\) & \(.070\) & \(.077\) \\
		\noalign{\smallskip} \hline \noalign{\smallskip}
		\multirow[t]{6}{*}{{.75}} 
		&\multirow[t]{2}{*}{{1}} 
		&\(\mathbb{P}(a_x)\)				&   \(.014\) & \(.033\) & \(.045\) & \(.033\) & \(.049\) & \(.075\) & \(.110\) & \(.125\) & \(.134\) & \(.190\) & \(.202\) & \(.216\) \\
		& & \(\tilde{a}_{x}\)		&   \(.011\) & \(.014\) & \(.014\) & \(.099\) & \(.072\) & \(.071\) & \(.209\) & \(.199\) & \(.193\) & \(.311\) & \(.305\) & \(.305\) \\
		&\multirow[t]{2}{*}{{3}} 
		& \(\mathbb{P}(a_x)\)				&  \(.037\) 	& \(.030\) & \(.031\) & \(.102\) & \(.108\) & \(.105\) & \(.178\) & \(.176\) & \(.172\) & \(.217\) & \(.212\) & \(.225\) \\
		& & \(\tilde{a}_{x}\)		&   \(.007\) & \(.007\) & \(.009\) & \(.044\) & \(.043\) & \(.045\) & \(.087\) & \(.087\) & \(.094\) & \(.138\) & \(.140\) & \(.142\) \\
		&\multirow[t]{2}{*}{{\(\infty\)}} 
		& \(\mathbb{P}(a_x)\)				&  \(.042\)	& \(.038\) & \(.034\) & \(.115\) & \(.111\) & \(.104\) & \(.163\) & \(.160\) & \(.167\) & \(.199\) & \(.198\) & \(.178\) \\
		& & \(\tilde{a}_{x}\)		&   \(.005\) & \(.005\) & \(.006\) & \(.024\) & \(.025\) & \(.026\) & \(.050\) & \(.048\) & \(.051\) & \(.072\) & \(.073\) & \(.074\) \\
\hline\noalign{\smallskip} 
\end{tabular}
\begin{flushleft}\footnotesize{
\(\mathbb{P}(a_x)\)=probability of excess effort (see Eq. \ref{eq:micro1}),  \(\tilde{a}_x\)=average excess effort made by the agent (see Eq. \ref{eq:micro2}).
}\end{flushleft}
\caption{Probabilities of making excess effort and average excess effort made by the agent}
\label{tab:micro-effects}
\end{scriptsize}
\end{sidewaystable}

\paragraph{Results}  
The effect introduced in Sec. \ref{sec:microscopic-effect} suggests that the agent's excess effort and the fluctuation range of the \textcolor{black}{predicted set of feasible actions} are positively correlated. The results presented in Tab. \ref{tab:micro-effects} support this conjecture and provide the following insight into the related dynamics. First, we can observe that the average excess effort increases with the turbulence of the environment. For relatively stable environments, the average excess effort is almost negligible. However, as the turbulence increases to \(\sigma=0.65x^{\ast}\), the range of excess effort made by the agent increases to a range of \(0.072\) to \(0.316\). 

Second, the results indicate that the agent's excess effort is also affected by his access to information about the environment. More (less) access to information results in less (more) excess effort. Environmental turbulence appears to have a moderating effect: If organizations operate in medium to highly turbulent environments (\(\sigma=0.25x^{\ast}\) to \(0.65x^{\ast}\)) and information access is decreased from \(m=\infty\) to \(m=1\), the extent of excess effort increases at least (most) by a factor of \(2.8\) (\(5.6\)). For more stable environments, the strength of the effect decreases. In addition, we can observe that a variation in the principal's tendency for a global search appears not to have significant effects on the average excess effort.

The analysis of the probability of making an excess effort complements the analysis conducted so far. First, limited access to information about the environment decreases the probability of making an excess effort. This observation is driven by the incentives provided to the agent: Recall that the principal employs an incentive scheme to make sure that the agent takes the action that the principal intends. From Sec. \ref{sec:macrodynamics} we already know that the principal has to reset her hill-climbing procedure whenever the position from which she starts her search moves in the infeasible region (see also period \(t+h+4\) in Fig. \ref{fig:macroeffect}), and more (less) information about the environment results in a less (more) frequent reset of the principal's search procedure. After a reset, the principal provides the agent with incentives to take an action that is usually more distant from the upper boundary of the action space (see Fig. \ref{fig:microeffect}). The agent responds to the stimuli and makes less of an effort. Consequently, the less frequently the principal has to reset her hill-climbing-based search procedure, the higher the probability for the agent to make an excess effort and vice versa. This effect is particularly pronounced in stable to mid-stable environments (up to \(\sigma=0.45x^{\ast}\)). In very turbulent environments, the fluctuation in the predicted action space is relatively high, so that the variation in the environment offsets the effect.

Second, if the size of the local search space increases, the probability of making excess effort rises, too. Recall that the size of the local search space translates into stability margins that can determine the robustness of the principal's solution to inaccurate predictions of the action space (see Sec. \ref{sec:macrodynamics}). In particular, low (high) values of parameter \(1/\lambda\) result in more (less) frequent resets of the search procedure. Thus, if the principal's search procedure is robust, \textcolor{black}{i.e., there is a sufficiently high stability margin}, she provides the agent with incentives to perform actions that are close to the upper boundary of the action space. This finding is more pronounced for firms that operate in rather stable environments, as increasing environmental turbulence overrules the robustness of the principal's search procedure. 

\subsection{Discussion}
\label{sec:discussion}

Perhaps most importantly, we show that (generally) the agent does not make the optimal effort, which, in turn, means that the efficiency of the solution proposed by the hidden-action model decreases if the idealized assumptions that are included in the hidden-action model are relaxed. In addition, we show that if the principal and the agent are given the opportunity to compensate for limited \textcolor{black}{access to} information by learning and a large memory, the feasible solution is at least close to the optimal one. Recently, \cite{Leitner2020} and \cite{Reinwald2020,reinwald2021,reinwald2022} were also concerned with the assumptions included in the hidden-action model. Their analysis, however, focuses exclusively on the emerging patterns at the macroscopic level (in terms of the task performance), and they come to similar conclusions related to the robustness of the incentive mechanism. Thus, the approach in this paper substantially complements previous research, as we focus on the dynamics that emerge from limited \textcolor{black}{access to} information at the \textit{microscopic level} in the sphere of both the principal and the agent, to provide a theoretical underpinning of the macroscopic patterns. 

\paragraph{The Sisyphus effect}

For the sphere of the principal, we have identified the \textit{Sisyphus effect}, which leads to a reset of the principal's search procedure if she overshoots the target and searches outside of the feasible region for (too high) action levels as the basis for the incentive mechanism. In parts, this effect refers to the findings related to search behavior in \cite{Levinthal1997} and \cite{Yang2007} or results associated with the exploration-exploitation dilemma, e.g., in \cite{March1991}, \cite{Uotila2009}, and \cite{Berger-Tal2014}. Previous research argues that some balance between \textcolor{black}{a local and global search} is required and that \textcolor{black}{a global search} is instrumental in the early phases of the search process. In contrast, \textcolor{black}{a local search} appears to be more efficient in later stages, i.e., once the performance has already been improved. For the context of reinforcement learning, \cite{Yen2002} argue that a proper balance between \textcolor{black}{searching locally} and \textcolor{black}{globally} is particularly relevant in dynamic environments. We take this into account and extend previous research on \textcolor{black}{the exploration-exploitation dilemma} in the field of managerial science by dynamic and endogenous boundaries of the search space, and argue that the size of local search spaces can aid as a stability margin to assure the robustness of solutions, which appears to be particularly useful in turbulent environments. 

\paragraph{The agent's excess effort}

For the agent's domain, we identify the micro-level dynamics that lead to the \textit{excess effort}. In particular, we analyze the behavioral dynamics that limited \textcolor{black}{access to} information unfolds in the context of hidden-action problems and find that the probability of making an excess effort is in the principal's sphere of control. In contrast, the extent of excess effort is a function of the agent's information. Thus, in limited information contexts, the principal's control over the agent's behavior is limited to the probability only. Our findings relate to previous research on reciprocal behavior. Reciprocity is referred to as a type of behavior that rewards kind actions and punishes unkind ones \citep{Falk2006}. The literature on reciprocity between managers and subordinates has its roots in research on \enquote*{gift exchange}: Assume that a gift is given to others without any associated payments. Following norm-gift-exchange models, social norms indicate that the receiver of the gift would eventually repay the gift in one or the other (possibly indirect) way \citep{Duffy2014,Mauss2002}. \cite{Akerlof1982} argues that the gift on the organization's side can be interpreted as payment exceeding what employees could earn from job options outside the firm.  Then, reciprocity is driven by the norm that employees repay the \enquote*{gift} putting in more than what is regarded as the minimum standard in effort \citep[see also][]{Yellen1984,Cropanzano1997}.\footnote{For reviews on the different concepts of reciprocity, the reader is, amongst others, referred to \cite{Fehr2000}, \cite{Dufwenberg2004}, and \cite{Gobel2013}.} Research on reciprocity is in line with previous research on the perception of what is perceived as fair behavior \citep{Kahneman1986} and patterns in actual behavior that are not in line with the frequent assumption of pure self-interest \citep{Roth1995,Fehr1998}. Reciprocal behavior might play an essential role in sequential settings \citep{Rabin1993,Dufwenberg2004}, i.e. when beliefs about how others behave are formed and when these beliefs are updated. Recall, in the model presented here, the principal's and the agent's respective information changes as they learn. Then, whenever the principal were to assess a previous action taken by the agent, the agent's excess effort might be interpreted as friendly (or hostile) behavior. From the agent's perspective, contracts might be perceived as overly beneficial or unfair after he has experienced his utility. Consequently, the micro-level dynamics that might result in excess effort might substantially contribute to sequential reciprocal behavior \citep{Dufwenberg2004}. Remarkably, the principal and the agent included in our model behave selfishly, since they purely maximize their utilities. However, they do so with limited \textcolor{black}{access to} information, which, in turn, might lead to the false perception that their behavior is driven by reciprocity considerations \citep{Fehr2000}. 

\section{Conclusive remarks}
\label{sec:conclusion}

This paper proposes an agent-based model of the hidden action problem introduced in \cite{Holmstrom1979}. While Holmstr\"om's model is undoubtedly fundamental for the development of principal-agent theory and has substantially advanced microeconomic research, it also builds on idealized assumptions about human capabilities, such as cognitive abilities to collect, process, and make sense of all information that is required to make optimal decisions. The proposed agent-based model includes a principal and an agent who are limited in their information. We observe how the principal and the agent learn and -- in an evolutionary sense -- improve the solution (the incentive scheme) and the performance that emerges from the principal's and the agent's decisions. Our most important contribution -- and the central extension of previous literature on this matter -- is identifying the micro-level dynamics that drive the results. For the sphere of the principal, we observe the so-called Sisyphus effect, while we refer to the dynamics in the agent's domain as excess effort. 

We believe that the insights provided in this paper are of equally high interest for researchers and for practitioners: From the researchers' perspective, an evaluation of the impact of assumptions related to information access on the robustness of incentive mechanisms, the dynamics emerging from limited information access, and a deeper understanding of the effects that drive the dynamics, might be beneficial for the scientific domains of mechanism design and management control. Viewed through the lens of practitioners, we believe that insights into the driving forces of individual behavior are particularly helpful as they might facilitate a correct interpretation of patterns in employees' behavior that might appear to be driven by reciprocity considerations. 

Of course, our research is not without limitations. \textcolor{black}{Most importantly, translating Holmstr\"om's hidden-action model \citep{Holmstrom1979} into an agent-based model that can be \enquote*{solved} via numerical simulation requires specifying some of the functions in the model. In consequence, the formulation becomes less generic. For example, we limit our analysis to a linear sharing rule. Future research might analyse further forms of sharing rules, production functions, and/or utility functions in this context.} Moreover, in our model, the principal and the agent rationally assess the situation and do not take reciprocity considerations into account. However, including sequential reciprocal behavior might also be a promising avenue for future research. \textcolor{black}{Also, a future research should perform a sophisticated analysis of the emergent sharing rule.} In addition, we argue that a stability margin can help reduce the Sisyphus effect's severity. Finally, in future research, introducing a standardized measure for the robustness of incentive schemes \textcolor{black}{as well an analysis of optimal stability margins} might be particularly useful for the fields of mechanism design and management control. \textcolor{black}{We model the principal to be a utility maximizer, i.e., they always aim for the best choice in terms of the best incentive scheme. If they followed the concept of satisficing, i.e., if they aimed for \enquote*{good enough} choices, they could inform their choices of what is a \enquote*{good enough} incentive scheme by the analysis of optimal stability margins, thereby making sure that \enquote*{good enough} levels of effort on the agent's side can be perpetuated for a longer period of time.} 
\section*{Conflict of interest}
\noindent The authors declare that they have no conflict of interest.

\section*{Role of the funding source}
\noindent The funding source had \textit{no involvement} in study design, the analysis and interpretation of data, in the writing of the manuscript, and in the decision to submit the article for publication.

\bibliography{mybibfile}

\appendix
\newpage
\section{Solution to the hidden action problem introduced in \cite{Holmstrom1979}}
\label{app:a}

To solve the program formalized in Eqs. \ref{eq1:maximization} to \ref{eq1:ICC} , \cite{Holmstrom1979} suppresses \(\theta\) and regards \(x\) as random variable with distribution \(F(x,a)\) that is parameterized by \(a\): \(F(x,a)\) is this distribution of \(\theta\) induced on the \(x\) via the function \(x=x(a,\theta)\). 

A change in \(a\) has an effect on the distribution of \(x\) (\(F_a(x,a)>0\)). \(F(x,a)\) has a density function where \(f_a(x,a)\) and \(f_{aa}(x,a)\) are well defined for all \((x,a)\).\footnote{Subscript \(a\) denotes the partial derivative with respect to \(a\).} The program can be reformulated to
\begin{subequations}
\begin{align}
\max_{s(x),a} 	\quad 	& 	\int U_P\left(x-s\left(x\right)\right)f(x,a)dx \label{appEq:maximization}\\
\textrm{s.t.} 			\quad 	&	\int [V(s(x)) - G(a)]f(x,a)dx \geq\underline{U} \label{appEq:PC}\\
 					 		&	\int V(s(x))f_a(x,a)dx = G'(a) \label{appEq:ICC}
\end{align}
\end{subequations}
The multipliers for Eqs. (\ref{appEq:PC}) and (\ref{appEq:ICC}) are denoted by \(\lambda\) and \(\mu\), respectively. Point-wise optimization, then, leads to
\begin{equation}
\label{eq:app-solutiona} 
\frac{U_P^\prime(x-s(x))}{V^\prime(s(x))}=\lambda + \mu \cdot \frac{f_a(x,a)}{f(x,a)}~.
\end{equation} 
For risk-neutral principals, Eq. (\ref{eq:app-solutiona}) reduces to 
\begin{equation}
\label{eq:app-solutionb} 
\frac{1}{V^\prime(s(x))}=\lambda + \mu \cdot \frac{f_a(x,a)}{f(x,a)}~.
\end{equation} 

In the first best case (i.e., situations in which \(a\) can be observed by the principal), Eq. \ref{appEq:ICC} is not binding and \(\mu=0\), which indicates that paying the agent a fixed compensation is optimal. In the second-best case, \(a\) is not observable for the principal and there is an incentive problem, which is why the principal provides the agent with incentives to increase the level of \(a\). Then, Eq. \ref{appEq:ICC} is binding and \(\mu>0\). For a more detailed discussion, the reader is referred to \cite{Holmstrom1979} and \cite{Lambert2001}.  

\end{document}